\documentclass[10pt,twocolumn,superscriptaddress,
english,10pt,prl,
floatfix,aps
]{revtex4-2}
\usepackage[utf8]{inputenc}
\usepackage{amsmath}
\usepackage{amssymb}
\usepackage{graphicx}
\usepackage{xspace}
\usepackage{physics}
\usepackage{float}
\usepackage[backref=none,bookmarksnumbered=true,bookmarks=true,  	bookmarksopen=true,colorlinks=true,citecolor=blue,linkcolor=blue, 	anchorcolor=green,urlcolor=blue,unicode=false]{hyperref}
\usepackage{fancyvrb}
\usepackage{version} 

\makeatletter
\@ifundefined{textcolor}{}
{%
 \definecolor{BLACK}{gray}{0}
 \definecolor{WHITE}{gray}{1}
 \definecolor{RED}{rgb}{1,0,0}
 \definecolor{GREEN}{rgb}{0,1,0}
 \definecolor{BLUE}{rgb}{0,0,1}
 \definecolor{CYAN}{cmyk}{1,0,0,0}
 \definecolor{MAGENTA}{cmyk}{0,1,0,0}
 \definecolor{YELLOW}{cmyk}{0,0,1,0}
}

\usepackage{soul}
\usepackage{braket}
\usepackage[backref=none,
bookmarksnumbered=true,
bookmarks=true,
bookmarksopen=true,
colorlinks=true,
citecolor=blue,
linkcolor=blue,
anchorcolor=green,
urlcolor=blue,unicode=false]{hyperref}

\let\baraccent=\= 
\renewcommand{\=}[1]{\stackrel{#1}{=}} 

\newcommand{\didv}{\ensuremath{\mathrm{d}I/\mathrm{d}V}\xspace}

\newcommand{\Isw}{$I_{\text{sw}}$}
\newcommand{\Ire}{$I_{\text{re}}$}

\newcommand{\Gpd}{$G_{\textrm{PD}}$}

\makeatother

\usepackage{babel}

\begin{document}
\title{Diode effect in Josephson junctions with a single magnetic atom}
\author{Martina Trahms}
\affiliation{Fachbereich Physik, Freie Universit\"at Berlin, Arnimallee 14, 14195 Berlin, Germany}

\author{Larissa Melischek}
\affiliation{Dahlem Center for Complex Quantum Systems and Fachbereich Physik, Freie Universit\"at Berlin, 14195 Berlin, Germany}

\author{Jacob F. Steiner}
\affiliation{Dahlem Center for Complex Quantum Systems and Fachbereich Physik, Freie Universit\"at Berlin, 14195 Berlin, Germany}

\author{Bharti Mahendru}
\affiliation{Fachbereich Physik, Freie Universit\"at Berlin, Arnimallee 14, 14195 Berlin, Germany}

\author{Idan Tamir}
\affiliation{Fachbereich Physik, Freie Universit\"at Berlin, Arnimallee 14, 14195 Berlin, Germany}

\author{Nils Bogdanoff}
\affiliation{Fachbereich Physik, Freie Universit\"at Berlin, Arnimallee 14, 14195 Berlin, Germany}

\author{Olof Peters}
\affiliation{Fachbereich Physik, Freie Universit\"at Berlin, Arnimallee 14, 14195 Berlin, Germany}

\author{Ga\"el Reecht}
\affiliation{Fachbereich Physik, Freie Universit\"at Berlin, Arnimallee 14, 14195 Berlin, Germany}

\author{Clemens B. Winkelmann}
\affiliation{Université Grenoble Alpes, CNRS, Institut Neél, 25 Avenue des Martyrs, 38042 Grenoble, France}

\author{Felix von Oppen}
\affiliation{Dahlem Center for Complex Quantum Systems and Fachbereich Physik, Freie Universit\"at Berlin, 14195 Berlin, Germany}

\author{Katharina J. Franke}
\affiliation{Fachbereich Physik, Freie Universit\"at Berlin, Arnimallee 14, 14195 Berlin, Germany}


\date{\today}
\begin{abstract}
Current flow in electronic devices can be asymmetric with bias direction, a phenomenon underlying the utility of diodes and known as non-reciprocal charge transport. The promise of dissipationless electronics has recently stimulated the quest for superconducting diodes, and non-reciprocal superconducting devices have been realized in various non-centrosymmetric systems. Probing the ultimate limits of miniaturization, we have created atomic-scale Pb--Pb Josephson junctions in a scanning tunneling microscope. Pristine junctions stabilized by a single Pb atom exhibit hysteretic behavior, confirming the high quality of the junctions, but no asymmetry between the bias directions. 
Non-reciprocal supercurrents emerge when inserting a single magnetic atom into the junction, with the preferred direction depending on the atomic species. Aided by theoretical modelling, we trace the non-reciprocity 
to quasiparticle currents flowing via Yu-Shiba-Rusinov (YSR) states inside the superconducting energy gap. Our results open new avenues for creating atomic-scale Josephson diodes and tuning their properties through single-atom manipulation.

\end{abstract}

\maketitle

\section{Introduction}

Ever since the invention of semiconductor p-n junctions, currents asymmetric in the direction of the applied bias voltage have been central to the development of electronic devices \cite{Scaff1947,Shockley1949}. In p-n junctions, non-reciprocal charge transport emerges from the band misalignment at the interface, which breaks inversion symmetry. In the absence of abrupt material interfaces, non-reciprocal charge transport usually occurs when broken inversion symmetry (e.g., by an electric field or the Rashba effect) is accompanied by broken time-reversal symmetry (e.g., by an applied magnetic field) \cite{Tokura2018}. If the current flows perpendicular to crossed electric and magnetic fields, its magnitude depends on the direction, a phenomenon known as the magnetochiral effect \cite{Rikken2005}. 

Non-reciprocal charge transport is particularly appealing for superconducting devices. They can exhibit dissipationless supercurrent in one direction, while the reverse direction is resistive, allowing for essentially unlimited resistance ratios. Diode behavior has recently been realized in noncentrosymmetric low-dimensional superconductors \cite{Wakatsuki2017, Qin2017,Bauriedl2022} as well as in inversion-symmetry-breaking stacks of different superconductors \cite{Ando2020}, making use of the strong magnetochiral effect when spin-orbit coupling and superconducting gap are of  comparable magnitude. The need for a time-reversal-breaking external magnetic field can be avoided by including magnetic interlayers \cite{Narita2022}.  

Josephson junctions provide an alternative platform for diode-like behavior in superconductors, offering additional tunability and potentially interfacing with superconducting qubits. Several corresponding experiments have recently observed non-reciprocal behavior. Baumgartner et al.\ \cite{Baumgartner2021} used a proximity-coupled two-dimensional electron gas with strong spin-orbit interaction, Pal et al.\ \cite{Pal2021} observed diode-like behavior in superconducting junctions in proximity to a topological semimetal, and Diez-Merida et al. \cite{Diez2021} in twisted bilayer graphene. While these devices required external magnetic fields to induce the diode effect, Wu et al.\ \cite{Wu2022} demonstrated rectification in a NbSe$_2$/Nb$_3$Br$_8$/NbSe$_2$ junction without magnetic fields (see also \cite{Bocquillon2017}).

\begin{figure*}[htp]\centering
\includegraphics[width=0.85\linewidth]{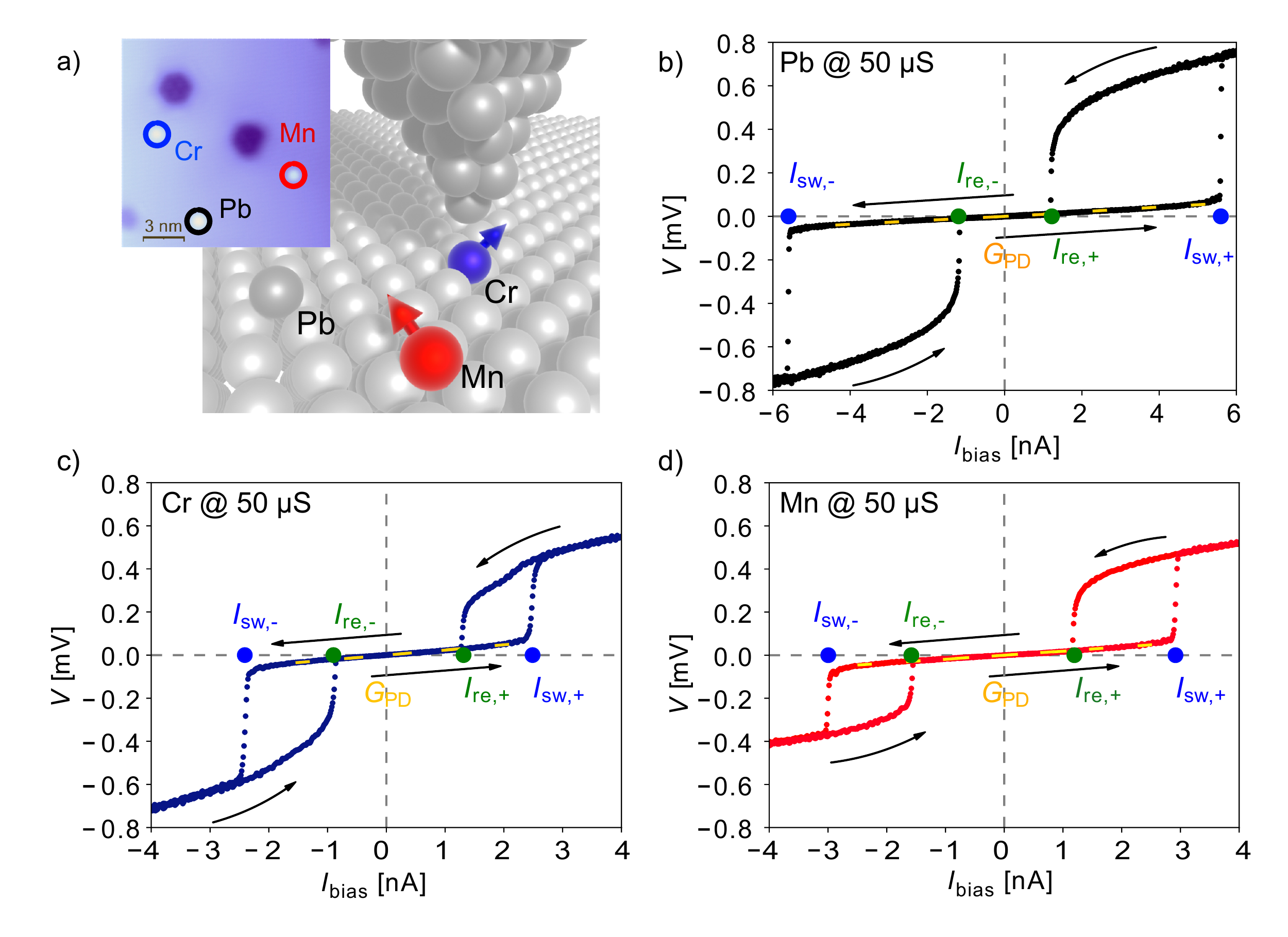}
\caption{\textbf{Single-atom Josephson junctions including Pb, Mn, and Cr atoms.}
(a) Sketch of STM-based Josephson junction including a single atom. Inset: STM topography of a Pb(111) surface with individual Pb, Mn, and Cr adatoms (colored circles); scanning parameters 50\ mV/ 50\ pA. (b-d) $V-I$ curves of current-biased Pb-Pb junctions including a (b) Pb, (c) Cr, and (d) Mn atom, measured at a normal-state conductance of $G_\mathrm{N}$\ = \ 50\ $\mu$S. Sweep directions are indicated by black arrows. Switching and retrapping events are indicated by blue and green dots, respectively. The slope at small currents (inverse of the phase-diffusion conductance \Gpd) 
is marked by a yellow dashed line.}
\label{Fig:fig1}
\end{figure*}

Here, we report that insertion of a single atom can induce diode-like behavior in Josephson junctions implemented using a scanning tunneling microscope (STM). Josephson coupling with and without adatoms has long been investigated using STMs with superconducting tips, focusing on spectroscopy of tunneling processes and excitations \cite{Naaman2001,Rodrigo2004,Bastiaans2019}, pair-density waves \cite{Hamidian2016,Liu2021}, phase diffusion \cite{Jaeck2017}, photon-assisted tunneling \cite{Roychowdhury2015,Kot2020,Peters2020}, Josephson spectroscopy \cite{Randeria2016, Kuester2021}, and $0-\pi$ transitions \cite{Karan2022}. While previous work on single-atom junctions focused on voltage-biased junctions, diode effects require current-biased measurements. We realize current-biased Josephson junctions and find diode-like behavior when including a single magnetic atom. We show that magnitude and sign of the diode effect can be tuned by the choice of atomic species. This makes our single-atom Josephson diodes a promising platform for studies of superconducting diodes, in particular when combined with single-atom manipulation to place the atoms into different configurations or to assemble them into nanostructures.

We also demonstrate that single-atom Josephson junctions constitute a versatile testbed for unraveling the physical mechanisms underlying the rectification. Our current-biased junctions exhibit a hysteretic voltage response, with the switching current (\Isw) -- marking the transition from dissipationless to resistive junction behavior upon increasing the current bias -- well separated from the retrapping current (\Ire) -- marking the reverse transition upon reducing the current. Unlike the critical current relevant in bulk superconductors, these properties are inherently related to the junction dynamics. In addition to the current-phase relation, the dynamics depends on the dissipative currents flowing in parallel to the supercurrent, the associated Johnson-Nyquist noise, and the junction capacitance \cite{Stewart1969, McCumber1968, Ambegaokar1969, Ivanchenko1969, Kautz1990}. While all of these can induce non-reciprocal behavior, we find that they affect the various characteristic currents in different ways. This allows us to identify a new mechanism which underlies the diode effect in our devices. 

\begin{figure*}[htp]\centering
\includegraphics[width=0.95\linewidth]{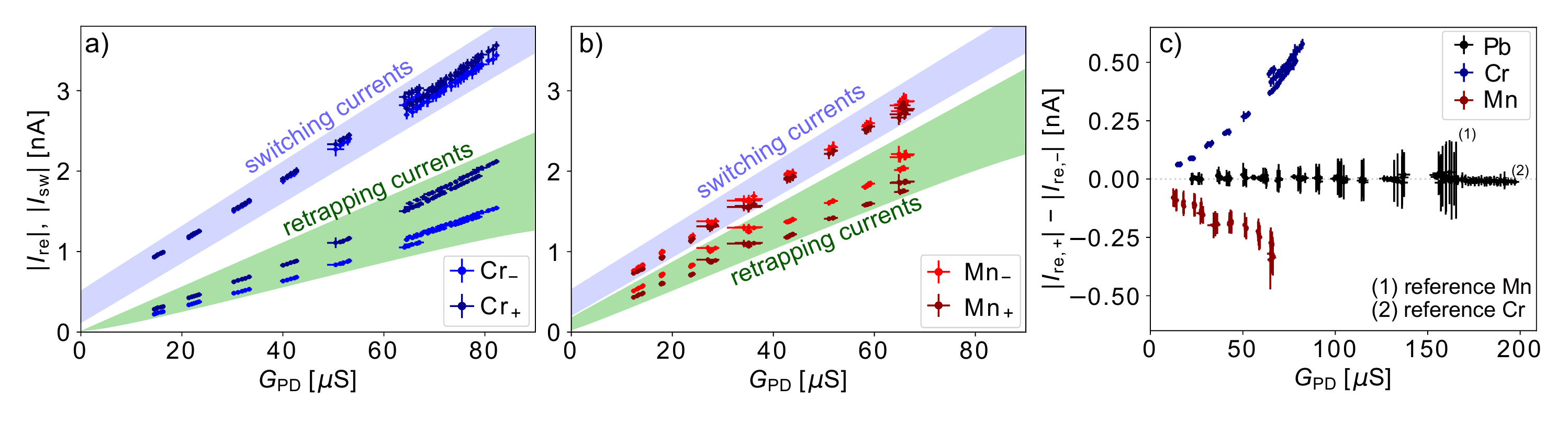}
\caption{\textbf{Non-reciprocity of switching and retrapping currents vs.\ junction transparency.} (a,b) Absolute values of retrapping and switching currents as extracted from $V-I$ curves for (a) Cr and (b) Mn junctions. Each data point averages over 100 sweeps recorded during longer measurement series started at normal-state conductances between (a) 25 and 50\ $\mu$S and (b) 20 and 50\ $\mu$S at 10\ mV. Error bars indicate the standard deviation of the average values. While piezoelectric creep slowly changes \Gpd\ (determined for each individual sweep), \Gpd\ remains essentially constant for the sweeps entering into a single data point (see Supplementary Note 2 for details). Positive/negative current bias is indicated by dark/bright colors and labeled as Cr$_+$/Cr$_-$ and Mn$_+$/Mn$_-$.  (c) Asymmetry $\Delta I_\mathrm{re}=|I_{\mathrm{re},+}|-| I_{\mathrm{re},-}|$ of the retrapping current for single-atom Cr, Mn, and Pb junctions. Pb junctions exhibit symmetric retrapping currents, while Cr and Mn atoms show non-reciprocities of opposite sign.}
\label{Fig:fig2}
\end{figure*}

\section{Current-biased single-atom Josephson junctions}

Figure \ref{Fig:fig1}a shows a sketch of our experimental setup. The Josephson junction is formed between the superconducting Pb tip of a STM and an atomically clean superconducting Pb(111) crystal with single Pb, Cr, and Mn atoms deposited on its surface (see STM image in Fig.\ \ref{Fig:fig1}a). Approaching these atoms by the tip allows us to investigate the influence of individual atoms on otherwise identical Josephson junctions. To establish these atomic-scale Josephson junctions, we approach the STM tip to the surface at a bias voltage well outside the superconducting gap, until a normal-state junction conductance of 50\ $\mu$S, of the order of but smaller than the conductance quantum, is reached. We then introduce a large resistor (1 M$\Omega$) in series with the junction, such that we effectively control the current bias of the junction. 

We first focus on junctions stabilized by a single Pb atom (Fig.\ \ref{Fig:fig1}b).  When
reducing the bias current from large positive currents, we observe a sharp reduction in the voltage drop across the junction at the retrapping current ($I_\mathrm{re}\approx 1.2$ nA). This marks the transition from resistive behavior dominated by quasiparticle tunneling (dissipative branch) to the near-dissipationless low-voltage state  dominated by Cooper-pair tunneling. Further reducing and eventually reversing the current bias to negative values, the junction abruptly transitions back to the dissipative branch at the switching current ($I_\mathrm{sw}\approx -5.6$ nA). Inverting the sweep direction of the current, the $V-I$ behavior exhibits a substantial hysteresis, but for pristine Pb-Pb junctions, the magnitudes of the switching and retrapping currents are independent of the bias direction (Fig.\ \ref{Fig:fig1}b). 

\section{Non-reciprocal Josephson currents induced by a single magnetic atom}

The Josephson junctions exhibit qualitatively different behavior, when the Pb atom is replaced by a single Cr or Mn atom (Fig.\ \ref{Fig:fig1}c,d). Incorporating a single magnetic atom into the junction significantly reduces the switching current compared to the pristine Pb junctions. This is consistent with a reduction of the Josephson peak in voltage-biased measurements on magnetic atoms and impurities \cite{Randeria2016, Kuester2021,Karan2022}. Intriguingly, we observe that the retrapping current and, to a much lesser extent, the switching current now depend on the direction of the current bias, so that the incorporation of a single magnetic atom makes the junction non-reciprocal. The behavior of our atomic-scale junctions differs qualitatively from observations of non-reciprocity in larger-scale junctions. While we observe the dominant asymmetry in the retrapping current, Refs.\ \cite{Pal2021, Wu2022} find stronger non-reciprocal behavior in the switching currents.   

To further investigate the non-reciprocity for the Cr and Mn junctions, we directly compare the switching and retrapping currents for both bias directions over a range of junction conductances (Fig.\ \ref{Fig:fig2}a,b). Accounting for the statistical nature of the switching and retrapping processes, every data point averages the switching or retrapping current over 100 current sweeps (see Supplementary Figs.\ 2-4 for full histograms). We quantify the junction conductance by the (inverse) slope of the $V-I$ curves in the low-voltage regime (cp.\ Fig.\ \ref{Fig:fig1}b-d), which we refer to as the phase-diffusion conductance \Gpd\ for reasons explained below. We find that the retrapping currents not only depend on the bias direction, but also on the particular type of magnetic atom. For Cr atoms the retrapping current is significantly larger in magnitude at positive bias ($I_{\mathrm{re},+}$) than at negative bias ($I_{\mathrm{re},-}$). For Mn atoms, the situation is just reversed. This is further illustrated in Fig.\ \ref{Fig:fig2}c, which shows the asymmetry $\Delta I_\mathrm{re}=|I_{\mathrm{re},+}| - |I_{\mathrm{re},-}|$ in the retrapping current as a function of \Gpd . The data suggest that, in addition, there is a considerably weaker asymmetry of the switching current (see also histograms in Supplementary Fig.\,1). 

\section{Phase dynamics}

\begin{figure*}[htp]\centering
\includegraphics[width=0.95\linewidth]{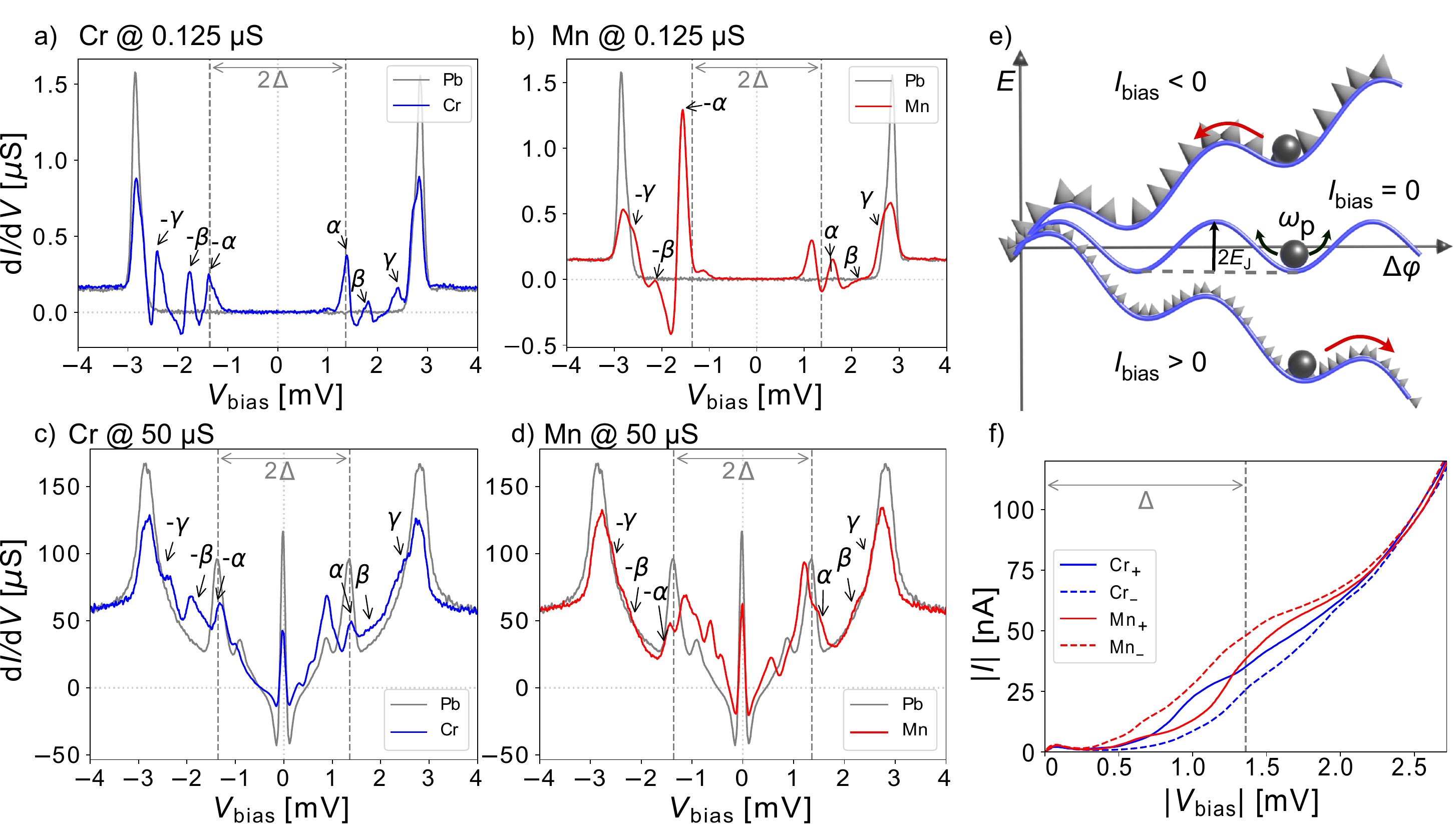}
\caption{\textbf{Yu-Shiba-Rusinov states as origin of non-reciprocity.} (a,b) Voltage-biased differential conductance spectra of (a) Cr and (b) Mn at a normal-state conductance of $G_\mathrm{N}=0.125 \mu S$. (Conductance set at 500 pA, 4 mV, lock-in modulation $V_\mathrm{rms} = 15 \mu $V.) The superconducting energy gap of the tip ($\Delta$) is marked by dashed lines. Reference spectra on Pb are shown in gray. YSR states are labeled as $\alpha$, $\beta$, $\gamma$. The YSR states are symmetric in energy about zero bias, but asymmetric in intensity due to electron-hole asymmetry. (c,d) \didv spectra of the same atoms as in (a,b) measured at  $G_\mathrm{N}=50 \mu$S. (Conductance set at 500 nA, 10mV, lock-in modulation $V_\mathrm{rms} = 15 \mu$V.) These spectra show a zero-bias Josephson peak and multiple Andreev reflections with and without exciting YSR states, in addition. As a result of the electron-hole asymmetry of the YSR states, the spectra exhibit intensities, which are distinctly asymmetric about zero bias. (e) Sketch of washboard potential (blue line) and friction (roughness of gray background) controlling the dynamics of a current-biased Josephson junction as represented by a phase particle (black sphere). The phase particle can be trapped in a minimum characterized by Josephson energy $E_\mathrm{J}$ and plasma frequency $\omega_p$ (trapped state) or slide down the washboard potential (running state). Non-reciprocal behavior originates with friction, which depends on bias direction as indicated by the different gray textures. (f) Current-voltage characteristics of voltage-biased Mn and Cr Josephson junctions for positive ($+$)/negative ($-$) voltages at $G_\mathrm{N}=50 \mu$S. The Cr junctions show a larger current magnitude at positive than at negative bias. The situation is opposite for Mn junctions. 
}
\label{Fig:fig3}
\end{figure*}

The hysteretic junction dynamics can be described within the model of a resistively and capacitively shunted Josephson junction (RCSJ). In this model \cite{Stewart1969, McCumber1968,Ambegaokar1969,Ivanchenko1969,Kautz1990}, the bias current $I_\mathrm{bias}$ applied to the junction splits between a capacitive current $I_c=C\dot V$, a dissipative current $I_d$ and its associated Nyquist noise $\delta I$, as well as the supercurrent $I_s(\varphi)$. Using the Josephson relation $V=\hbar \dot\varphi/2e$ for the voltage $V$ across the junction and assuming Ohmic dissipation, $I_d=V/R$, the superconducting phase difference $\varphi$ across the junction can be described as a Brownian particle moving in a tilted washboard potential (Fig.\ \ref{Fig:fig3}e),
\begin{equation}
    (\hbar C/2e)\ddot \varphi + (\hbar /2eR)\dot \varphi + I_s(\varphi) +\delta I = I_\mathrm{bias}.
    \label{Eq:RCSJ}
\end{equation}
The tilted washboard potential subjects the Brownian particle to a constant force associated with the bias current $I_\mathrm{bias}$, in addition to a periodic force originating from the current-phase relation $I_s(\varphi)$ of the junction.

Focusing first on the case of pristine Pb junctions, the hysteretic behavior emerges as follows. At small bias currents, the phase is trapped in a minimum of the tilted washboard potential, corresponding to supercurrent flow. Increasing the bias current tilts the washboard potential and lowers the potential barrier for activation of the phase particle out of the minimum. Once the phase particle escapes, it crosses over to a running solution associated with a voltage drop across the junction (switching current). Conversely, when reducing the bias current, inertia makes the phase particle retrap into a minimum only at a smaller current, at which friction balances the energy gained due to the tilt of the washboard potential (retrapping current). In our junctions, switching occurs long before the bias current reaches the critical current $I_c$ (estimated to be $107$\,nA based on the Ambegaokar-Baratoff formula \cite{Ambegaokar1963}), at which the tilted washboard potential loses its minima, indicating the importance of the Nyquist noise $\delta I$. We note that we observe a small voltage drop also in the nominally trapped state at small bias current. This behavior is familiar for small junctions and a well-understood consequence of frequency-dependent damping \cite{Kautz1990}, leading to residual phase diffusion and the zero-bias conductance \Gpd\ (see also Supplementary Note 5).

While this basic RCSJ model cannot describe dynamics of Josephson junctions, which is asymmetric in the bias directions, several extensions of the RCSJ model are known to predict non-reciprocal behavior. Diode-like behavior can originate with an asymmetric current-phase relation \cite{Yokoyama2013, Dolcini2015, Chen2018, Baumgartner2021, Pal2021, Davydova2022, Souto2022} or non-linear corrections to the capacitive term associated with the quantum capacitance \cite{Misaki2021}. An asymmetric current-phase relation implies a non-reciprocal switching current, inconsistent with our observations. Non-linear corrections to the capacitive term induce asymmetric retrapping currents. However, this requires a junction with strongly asymmetric carrier densities on its two sides, a feature that is absent for our Pb-Pb junctions. 

\section{Origin of non-reciprocity}

Non-reciprocity of the retrapping current, coexisting with less asymmetric switching currents, suggests instead that the non-reciprocity originates with the damping properties of the junction. Microscopically, the dissipative current $I_d$ accounts for the quasiparticle current flowing in parallel to the supercurrent as well as dissipation into the electromagnetic environment. While the effects of the electromagnetic environment are expected to be independent of the bias direction, the quasiparticle current can be non-reciprocal. 

The asymmetry of the quasiparticle current is directly accessible in voltage-biased measurements, with a superconducting tip, of the same junctions. Figure \ref{Fig:fig3}a,b shows tunneling spectra on Cr and Mn atoms at small junction conductance (0.125 $\mu$S), revealing strong subgap resonances of the differential conductance \didv (and thus current). In addition to the coherence peaks at (2.72 $\pm$ 0.05) mV, we resolve three pairs of conductance peaks, labeled by $(\alpha,\beta,\gamma)$, which we identify with Yu-Shiba-Rusinov (YSR) states \cite{Ruby2016} within the superconducting energy gap $\Delta$. Peaks occurring at voltages $e|V|<\Delta$ originate from the same states, albeit probed by thermally excited quasiparticles \cite{Ruby2015}. While the YSR resonances must occur symmetrically in energy, they need not have symmetric intensities \cite{Yazdani1997, Ruby2016, Choi2017}. We observe that this asymmetry is particularly pronounced for the deepest ($\alpha$) YSR state of Mn. By comparison, Cr exhibits weaker, but still well-resolved asymmetries of the YSR-state intensities. Importantly, there is no asymmetry in the corresponding \didv traces for the junction stabilized on a Pb atom (see gray traces in Fig.\ \ref{Fig:fig3}a,b). 
  
These results indicate that the asymmetric subgap conductance associated with the YSR resonances is a natural source of the observed non-reciprocal behavior. However, the spectra in Fig.\ \ref{Fig:fig3}a,b were taken in the weak-tunneling regime, where the YSR resonances are well resolved, and are thus not of immediate relevance to the Josephson-junction regime at stronger tunneling. Figure \ref{Fig:fig3}c,d shows \didv spectra at junction conductances corresponding to the Josephson-junction regime. For the pristine Pb junctions, the larger junction conductance enables additional transport processes inside the gap due to Cooper-pair tunneling at zero bias (Josephson peak) and multiple Andreev reflections above the threshold voltages of $eV=2\Delta/n$ with $n=2,3,...$ (Fig.\ \ref{Fig:fig3}c,d, gray traces). Consistent with the weak-tunneling case, the \didv traces of pristine Pb junctions remain independent of bias direction at high junction conductance. 
 
For the Cr and Mn junctions at higher junction conductance, we observe an even richer in-gap structure, with intensities which are clearly asymmetric in the bias directions. We attribute the additional features to multiple Andreev processes exciting a YSR state of energy $\varepsilon$ in addition to quasiparticles in the electrodes. These processes have threshold energies of $eV=(\Delta +\varepsilon)/n$ \cite{Randeria2016, Farinacci2018} and reflect the asymmetry of the underlying YSR states. The resulting asymmetry in the subgap current is shown in Fig.\ \ref{Fig:fig3}f. Importantly, the quasiparticle current for Cr is larger at positive bias voltages. Since a larger quasiparticle current implies stronger dissipation, this is consistent with the larger retrapping current for this bias direction of the current-biased Josephson junction. The situation is just reversed for Mn, again consistent with the asymmetry of the retrapping current. 

\begin{figure}[htp]\centering
\includegraphics[width=0.95\linewidth]{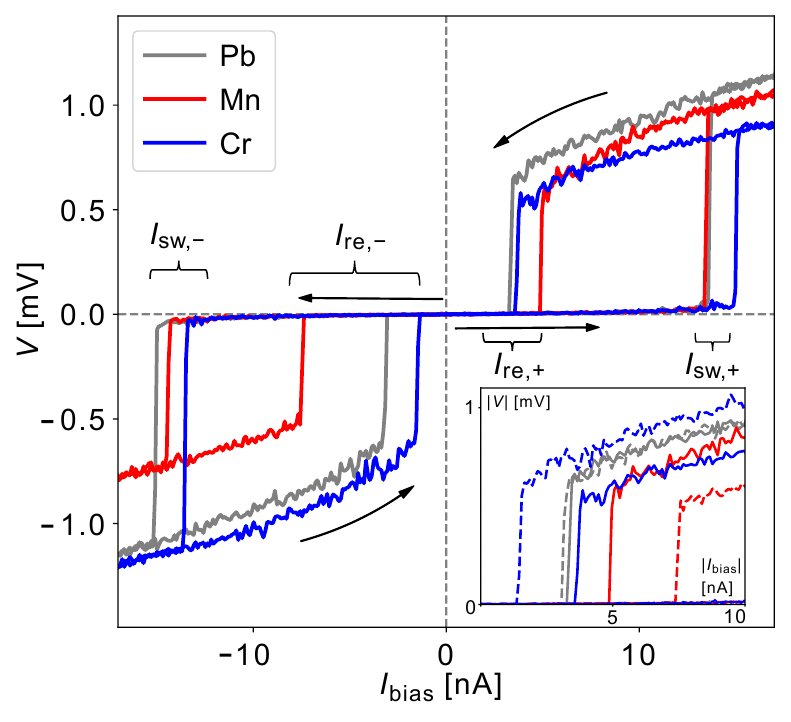}
\caption{\textbf{Modeling of non-reciprocity within RCSJ model.} 
Simulated hysteretic $V-I$ traces based on the extended RCSJ model, accounting for asymmetric as well as frequency-dependent friction. Traces for Pb (gray), Cr (blue), and Mn (red) use the non-Ohmic quasiparticle current $I_d(V)$ as extracted from the corresponding experimental data in Fig.\ \ref{Fig:fig3}c,d as input. All other model parameters (see Supplementary Note 5) are identical to highlight the effect of asymmetric friction. Inset: $|V|-|I|$ traces over a smaller range of bias currents using the same data as in the main panel to bring out the asymmetry, showing only the retrapping currents. Full (dashed) lines correspond to positive (negative) current bias with color coding as in main panel.   
}
\label{Fig:fig4}
\end{figure}

To further corroborate that an asymmetric quasiparticle current can induce non-reciprocal behavior of Josephson junctions, we perform numerical simulations for an extended RCSJ model \cite{Kautz1990}. We include frequency-dependent damping, allow for a non-linear and asymmetric dissipative $I_d(V)$, and account for the Johnson-Nyquist noise associated with the damping. To isolate the effect of asymmetric damping, we extract $I_d(V)$ based on the experimental data in Fig.\ \ref{Fig:fig3}f for the Pb, Cr, and Mn junctions, but otherwise use identical model parameters (for details, see Supplementary Note 5). Figure \ref{Fig:fig4} shows representative $V-I$ traces, which are symmetric for Pb, but exhibit asymmetric retrapping currents for Cr and Mn. The asymmetries clearly reproduce the sign found in experiment (cp., Fig.\ \ref{Fig:fig1}b-d). Consistent with experiment, our simulations also reproduce a weak asymmetry in the switching current (see histograms and discussion in Supplementary Note 5). We finally comment on the symmetry conditions for non-reciprocity originating from quasiparticle damping. Inversion symmetry is explicitly broken by single-atom junctions with the adatom attached to one of the electrodes. At the same time, the junction is time-reversal symmetric since in the absence of an external magnetic field, the spin of the magnetic molecule remains unpolarized. Instead, the asymmetric weights of the YSR resonances and hence the non-reciprocity are a consequence of broken particle-hole symmetry. 

\section{Conclusions}

Developing device applications for Josephson diodes requires a thorough understanding of the mechanisms underlying their non-reciprocity. Probing the limits of miniaturization, we have created and investigated Josephson diodes whose asymmetry is induced by the presence of a single magnetic atom within the junction. The single-atom nature of our junctions admits a comprehensive understanding of the observed non-reciprocity, and we find that its origin is qualitatively different from that underlying observations in larger-scale devices. We trace the non-reciprocity of our junctions to dissipation induced by quasiparticle currents flowing in parallel to the supercurrent. In the presence of magnetic atoms, the quasiparticle current can flow via YSR subgap resonances, which become asymmetric in the bias direction when particle-hole symmetry is broken. At the relevant junction conductances, the quasiparticle current involves not only direct single-electron tunneling into the YSR states, but also multiple Andreev reflections exciting the subgap states and thereby contributing to the asymmetry of the quasiparticle current. 

Our atomic-scale Josephson junctions provide exquisite flexibility for tuning the non-reciprocal behavior. We have already shown that the magnitude of the asymmetry can be tuned via the junction conductance and that the sign of the asymmetry depends on the atomic species inserted into the junction. Considerable opportunities are opened by combining atomic-scale Josephson junctions with single-atom manipulation. The asymmetry is expected to depend sensitively on the adsorption site of the magnetic atom and can be manipulated by bottom-up creation of atomic assemblies. Thus, our results pave the way towards designing Josephson diodes with a large degree of functional flexibility.  

\section{Methods}

The Pb(111) crystal was cleaned by multiple cycles of Ne$^{+}$ sputtering and subsequent annealing under ultra-high vacuum conditions.  Using an electron beam evaporator, magnetic adatoms (Chromium and Manganese) were deposited on the clean substrate held at 30 K. The as-prepared sample was then investigated in a Crea\-Tec STM at 1.3 K. The tungsten tip was coated with a sufficiently thick layer of Pb by dipping it into the crystal surface until a full superconducting gap is observed ($\Delta_\mathrm{tip}=\Delta_\mathrm{sample}$). Differential-conductance spectra at large junction resistance reveal the quality of the superconductor--superconductor junction by a superconducting gap of size $2\Delta_\mathrm{tip}+2\Delta_\mathrm{sample}=4\Delta$ around the Fermi level, flanked by a pair of coherence peaks \cite{RubyPb15} (gray spectra in Fig.\ \ref{Fig:fig3}a,b). 

As Josephson spectroscopy is performed at junction conductances of 20 $\mu$S or higher, exceptional tip stability is required to withstand the forces acting at these conductances. Smaller indentations are performed to improve the stability and sharpness of the tip. Individual Pb atoms from the tip apex were deposited by controlled approaches to the flat surface. Measurements were then done on individual Pb, Cr, or Mn adatoms on the Pb(111) surface. The Cr and Mn atoms were pulled out from the initial adsorption site by approaching with the STM tip \cite{Ruby2016}. 

Josephson spectroscopy was performed by increasing the current set point at a constant bias voltage of 10 mV until reaching the desired junction conductance. After tip stabilization, a large series resistor $R_\mathrm{series}$ = 1 M$\Omega$ was introduced into the bias line. This resistance is sufficiently large compared to the junction resistances, so that the junction is effectively current biased. Current-biased Josephson spectroscopy was then performed by sweeping the current bias back and forth between positive and negative values at a rate of 100 nA/s to 320 nA/s. Positive current corresponds to tunneling of electrons from tip to sample. For statistical analysis, we perform between 500 and 2000 sweeps in each direction. The STM feedback was turned off during the time of the measurement. 

The data analysis was performed using a dedicated Python program. Switching and retrapping events were extracted by taking the derivative of the individual $V-I$ curves, which were previously smoothed by a standard Gaussian routine. We also determined \Gpd\ from the slope of the $V-I$ curve in the trapped state. In analyzing the data, we account for a number of instrumental effects. (i) A slow creep of the piezoelectric elements causes the tip to drift towards the surface, effectively changing the junction conductance. We continuously monitor \Gpd\ to characterize the junction and plot all switching and retrapping currents vs.\ \Gpd. (ii) The differential amplifier used during the Josephson measurements introduces a slowly shifting voltage offset, that we subtract from the individual $V-I$ curves. (iii) The voltage/current source has a small offset. For this reason, we correct the entire data set, including the data measured on the magnetic adatoms, by the mean offset for all recorded data on the pristine Pb-Pb junction under identical measurement conditions, i.e., identical tip and tip locations. (iv) At high junction conductances, the voltage drop across the series resistance of the external circuit becomes non-negligible in the voltage-biased measurements. We correct for this by
calibrating the voltage to the superconducting gap size of the Pb-Pb junction.

\section*{Acknowledgements}
We thank K.\ Biel for assistance in preliminary measurements and C.\ Lotze for general technical support. We acknowledge financial support by the Deutsche Forschungsgemeinschaft (DFG, German Research Foundation) through projects 277101999 (CRC 183, project C03), FR2726/5, and SFB 910 (project A11), as well as by Agence Nationale de la Recherche under grant JOSPEC.

\bibliographystyle{naturemag}

\clearpage

\renewcommand{\figurename}{Supplementary Figure}
\renewcommand{\figurename}{Supplementary Figure}
\renewcommand{\tablename}{Supplementary Table}

\setcounter{figure}{0}
\setcounter{section}{0}
\setcounter{equation}{0}
\setcounter{table}{0}

\onecolumngrid

\newcommand{\vsigma}{\mbox{\boldmath $\sigma$}}

\section*{\Large{Supplementary Material}}

\maketitle

\section{Supplementary Note 1: Statistics of switching and retrapping currents}

\begin{figure*}[htp]\centering
\includegraphics[width=0.95\linewidth]{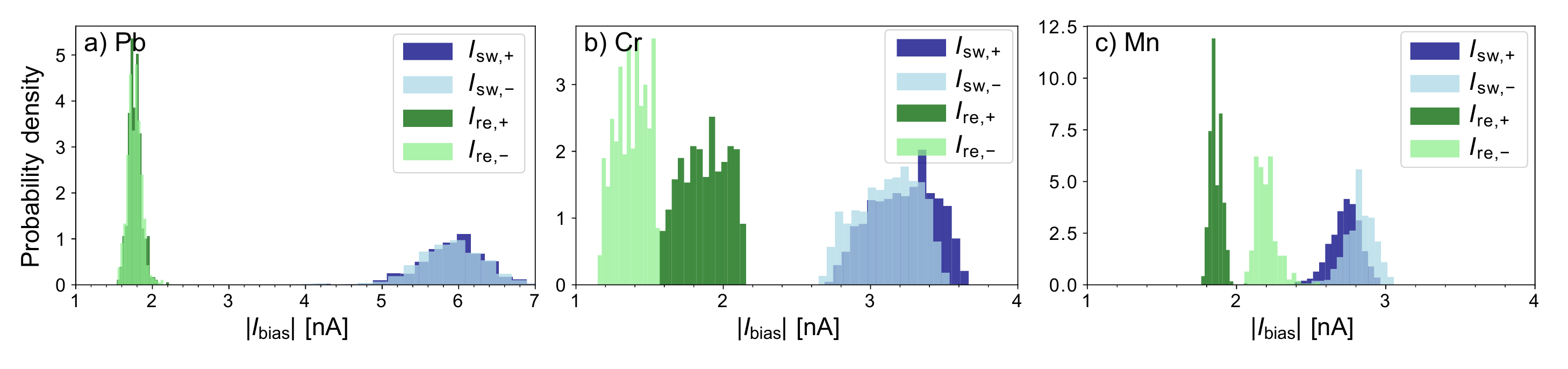}
\caption{\textbf{Statistics of switching and retrapping currents for single-atom Josephson junctions.} (a-c) Histograms of absolute values of switching and retrapping currents for the two bias directions, as extracted from individual $V-I$ curves for (a) Pb, (b) Cr, and (c) Mn junctions. The histograms in a) and c) include data extracted from 500 sweeps and b) includes 2000 sweeps for each current direction. The junction conductances $G_\mathrm{N}$ were set at 10\ mV to 50 $\mu$S. The distributions of switching and retrapping currents arise from the stochastic nature of switching and retrapping events, and are further broadened by piezoelectric creep while taking the 500 to 2000 sweeps (see Supplementary Note 2 for histograms without this additional broadening). 
}
\label{Fig:Sfig1}
\end{figure*}

As described in the main text, we create a Josephson junction by approaching the STM tip to the surface at a bias voltage (10 mV) far above the superconducting energy gap until the desired normal-state junction conductance (few tens of $\mu$S) is reached. We effectively current bias the junction by inserting a large series resistor ($R_\mathrm{series} = 1$ M$\Omega$) into the bias line and sweep the current (a few nA) in both directions. The transition from the resistive to the low-resistance state (\Ire) is seen as a sudden drop in the voltage, while switching from the low-dissipation to the dissipative branch (\Isw) occurs as a sudden increase in the voltage. Both events are stochastic in nature, due to Johnson-Nyquist noise. For this reason, we complement single sweeps by histograms of switching and retrapping currents extracted from a larger set of $V-I$ curves. Non-reciprocity of the switching and retrapping current is then seen as asymmetries between the histograms for positive and negative bias. Supplementary Fig.\ \ref{Fig:Sfig1} shows corresponding histograms extracted from 500 to 2000 sweeps recorded on Pb, Cr, and Mn junctions with $G_\mathrm{N}$ equal to 50 $\mu$S. For the Pb junction, the histograms of the switching currents $|I_\mathrm{sw,+}|$ and $|I_\mathrm{sw,-}|$ exhibit broad Gaussian-like distributions, with the same average (($5.9 \pm 0.4$) nA) for both bias directions. The histograms of the retrapping currents $|I_\mathrm{re,+}|$ and $|I_\mathrm{re,-}|$ are more narrow (($1.8 \pm 0.1$) nA), but also independent of bias direction. The histograms for Cr and Mn junctions are qualitatively different. The histograms of the retrapping currents exhibit a clear relative shift between the two bias directions, leading to different absolute values of the averages of $I_\mathrm{re,+}$ [($1.9 \pm 0.2$) nA for Cr and ($1.86 \pm 0.04$) nA for Mn] and $I_\mathrm{re,-}$ [($-1.4 \pm 0.2$) nA for Cr and ($-2.18 \pm 0.06$) nA for Mn].
The histograms of the switching current exhibit a noticeable, but much weaker dependence on the bias direction. 

\section{Supplementary Note 2: Analysis of switching and retrapping currents as a function of $G_\mathrm{\textbf{PD}}$}

\begin{figure*}[htp]\centering
\includegraphics[width=0.8\linewidth]{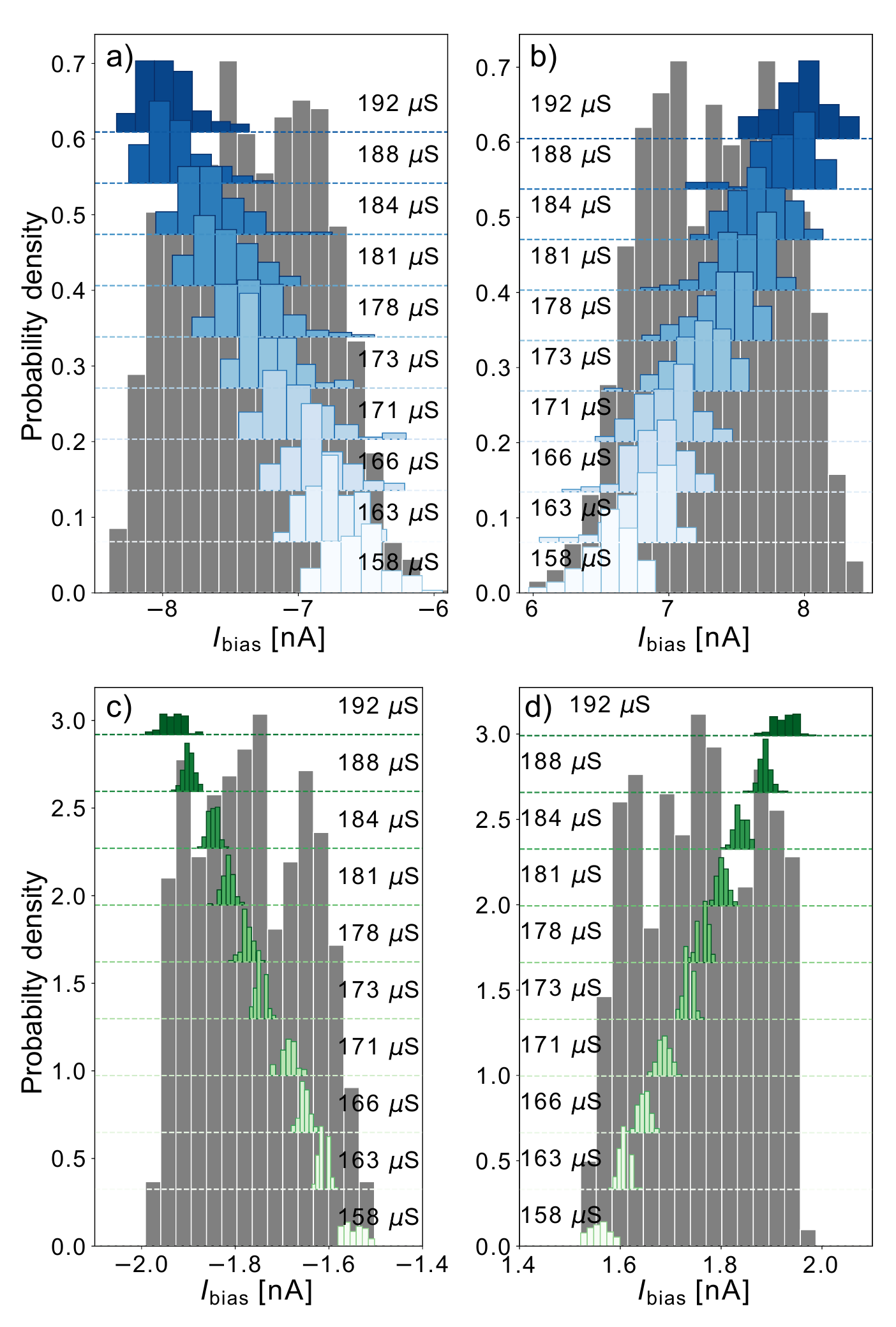}
\caption{\textbf{Evolution of histograms with \Gpd\ for a Pb junction.} The gray histograms (background) are extracted for (a,b) switching and (c,d) retrapping currents from 2000 individual $V-I$ curves, recorded after setting the junction to a (high-voltage) conductance of $G_\mathrm{N}=50$ $\mu$S. The (a) positive-bias and (b) negative-bias switching currents were divided into bins of 100 sweeps each (blue histograms). The same procedure was implemented for (c) positive-bias and (d) negative-bias retrapping currents (green histograms). Every other histogram is omitted for clarity. \Gpd\  varies due to piezoelectric drift. Its average value is indicated for each of the histograms. The piezoelectric drift to larger \Gpd\ over the course of the measurement is reflected in shifts to higher absolute values of switching and retrapping currents. Note that these data were recorded with a different tip than those in Supplementary Fig. \ref{Fig:Sfig1}.
}
\label{Fig:Sfig2_Pb}
\end{figure*}

\begin{figure*}[htp]\centering
\includegraphics[width=0.8\linewidth]{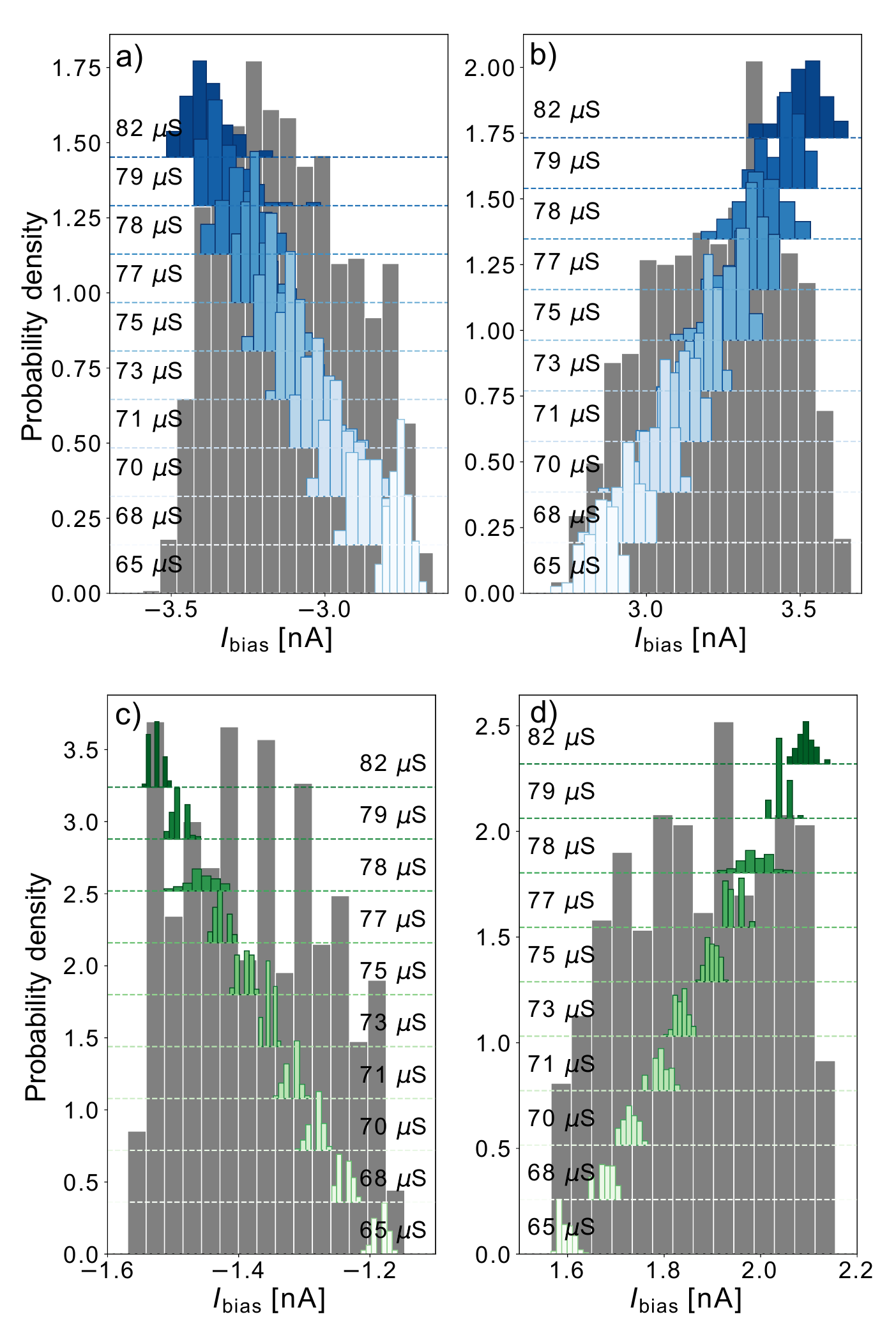}
\caption{\textbf{Evolution of histograms with \Gpd\ for a Cr junction.} The gray histograms (background) are extracted for (a,b) switching and (c,d) retrapping currents from 2000 individual $V-I$ curves, recorded after setting the junction to a (high-voltage) conductance of $G_\mathrm{n}=50$ $\mu$S. The (a) positive-bias and (b) negative-bias switching currents were divided into bins of 100 sweeps each (blue histograms). The same procedure was implemented for (c) positive-bias and (d) negative-bias retrapping currents (green histograms). Every other histogram is omitted for clarity. \Gpd\ varies due to piezoelectric drift. Its average value is indicated for each of the histograms. The piezoelectric drift to larger \Gpd\ over the course of the measurement is reflected in shifts to higher absolute values of switching and retrapping currents. 
}
\label{Fig:Sfig2_Cr}
\end{figure*}

\begin{figure*}[htp]\centering
\includegraphics[width=0.8\linewidth]{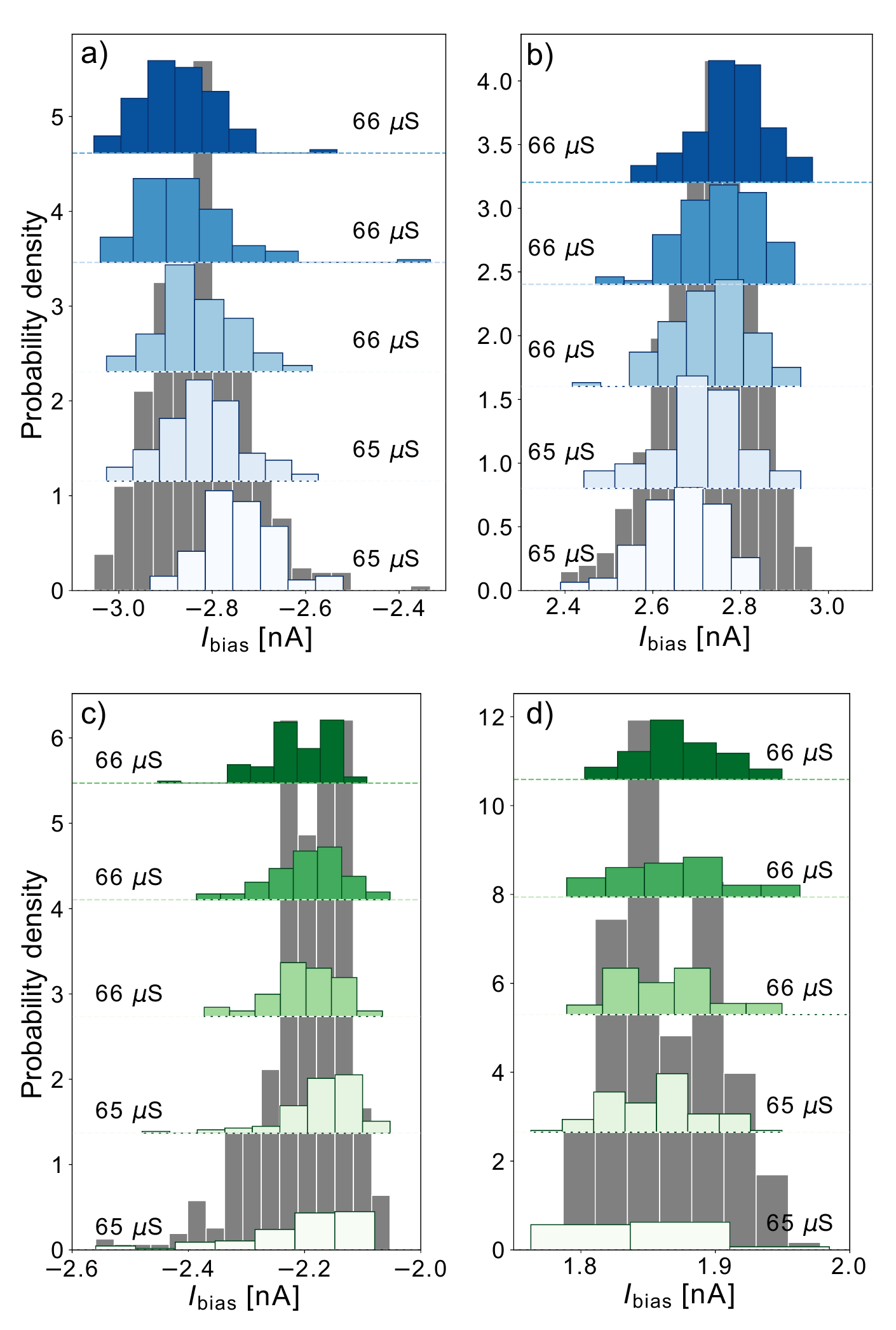}
\caption{\textbf{Evolution of histograms with \Gpd\ for a Mn junction.} The gray histograms (background) are extracted for (a,b) switching and (c,d) retrapping currents from 500 individual $V-I$ curves, recorded after setting the junction to a (high-voltage) conductance of $G_\mathrm{n}=50$ $\mu$S. The (a) positive-bias and (b) negative-bias switching currents were divided into bins of 100 sweeps each (blue histograms). The same procedure was implemented for (c) positive-bias and (d) negative-bias retrapping currents (green histograms). \Gpd\ varies due to piezoelectric drift. Its average value is indicated for each of the histograms. The piezoelectric drift to larger \Gpd\ as well as the shifts to higher absolute values of switching and retrapping currents are less pronounced than in Supplementary Fig.\ \ref{Fig:Sfig2_Pb}, since the time of measurement was much shorter. 
}
\label{Fig:Sfig2_Mn}
\end{figure*}

The histograms in Supplementary Fig.\ \ref{Fig:Sfig1} reflect the stochastic nature of the switching and retrapping processes, but are further broadened by piezoelectric creep over the course of the measurement. The creep effectively increases the junction conductance (as quantified by the phase-diffusion conductance \Gpd) with time. We minimize the creep-induced broadening by using each 100 consecutive sweeps for separate histograms with an associated average \Gpd. Supplementary Figs.\ \ref{Fig:Sfig2_Pb}-\ref{Fig:Sfig2_Mn} illustrate this analysis. The histogram for the earliest 100 sweeps are shown at the bottom of each panel. Histograms obtained from subsequent batches of 100 sweeps correspond to larger junction conductances \Gpd, as indicated in the figure. This increase is accompanied by an increase in $|I_\mathrm{sw}|$ and $|I_\mathrm{re}|$ as seen by a shift of the corresponding histograms. This scheme is the basis for Fig.\ 2 of the main text, which collects the average retrapping currents along with the standard deviations of all of these histograms.

\section{Supplementary Note 3: Comparison of switching currents}

\begin{figure*}[htp]\centering
\includegraphics[width=0.95\linewidth]{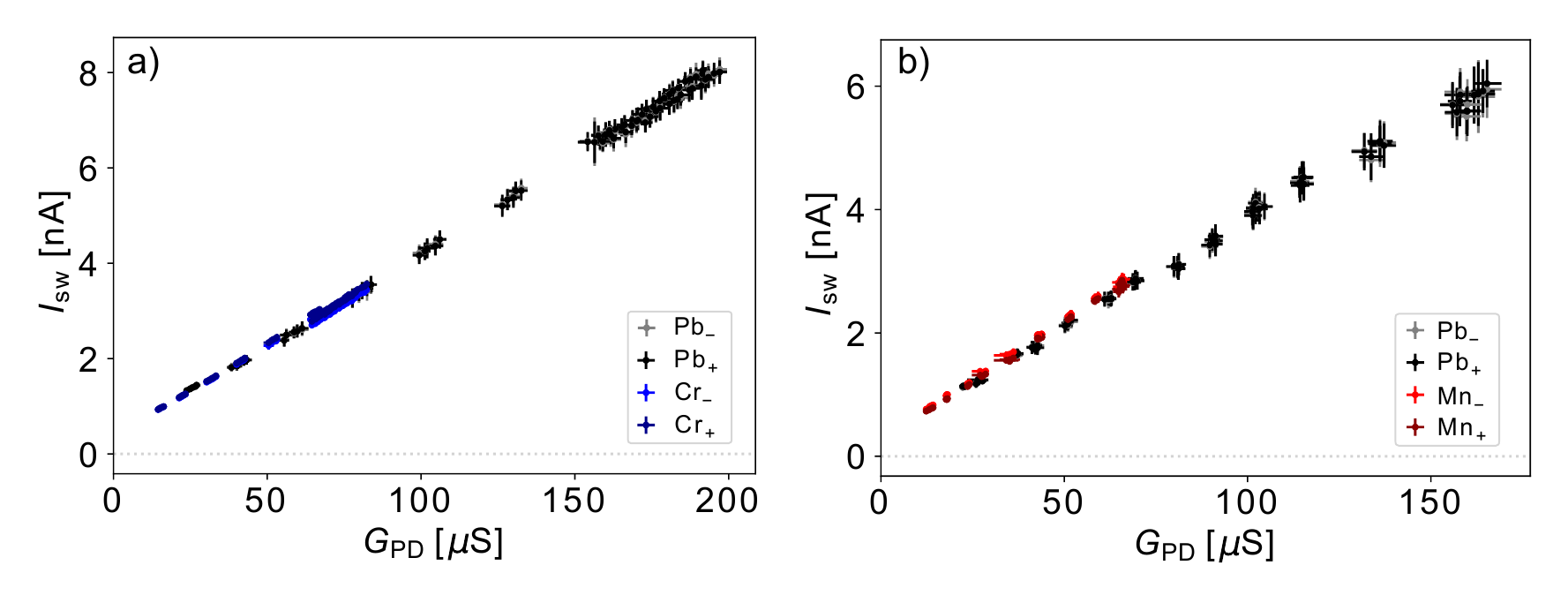}
\caption{\textbf{Comparison of switching currents for the Cr and Mn junctions with the reference data for Pb junctions.} (a) Extracted positive and negative switching currents for Cr and Pb junctions with normal state conductances $G_\mathrm{N}$ between 20 and 50 $\mu$S. The data were acquired with the same tip and under similar measurement conditions. (b) Extracted positive and negative switching currents for Mn and Pb junctions with normal state conductances $G_\mathrm{N}$ between 20 and 50 $\mu$S . The switching current depends linearly on \Gpd, with the same slope for magnetic and non-magnetic atoms provided data are taken under corresponding measurement conditions.
}
\label{Fig:Sfig3}
\end{figure*}

Supplementary Fig.\ \ref{Fig:Sfig3} shows the switching currents of (a) Cr and (b) Mn junctions as a function of \Gpd, in both cases compared to Pb junctions measured with the same tip. For identical tips, the switching currents $|I_\mathrm{sw}|$ show almost the same linear dependence on \Gpd, justifying the use of \Gpd\ as a suitable measure of junction conductance. 

\section{Supplementary Note 4: Influence of the STM tip}

\begin{figure*}[htp]\centering
\includegraphics[width=0.95\linewidth]{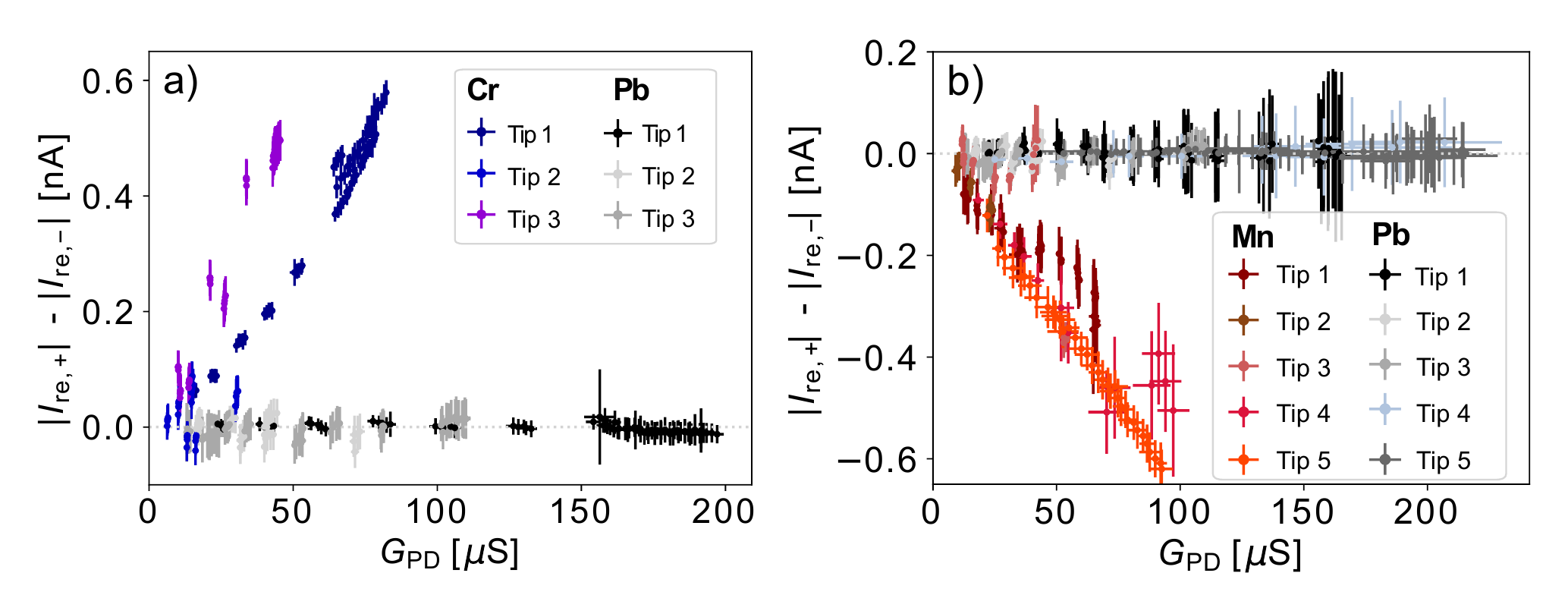}
\caption{\textbf{Influence of tip shape.} Non-reciprocity of the retrapping currents ($\Delta I_\mathrm{re}=|I_\mathrm{re,+}| -| I_\mathrm{re,-}|$) for (a) Cr and Pb and (b) Mn and Pb junctions as measured with different tips. Pb junctions have symmetric retrapping currents, while Cr and Mn junctions show non-reciprocity of the retrapping current.  
The precise value of the asymmetry varies between different tips, but the sign of the asymmetry is consistently opposite for Cr and Mn. The (high-voltage) junction conductances $G_\mathrm{N}$ were set between 20 and 50 $\mu$S at 10 mV. \Gpd\ was determined from individual $V-I$ sweeps as described in Supplementary Note 2. The asymmetry was derived from \Isw\ and \Ire\ after averaging over 100 sweeps.  }
\label{Fig:Sfig4}
\end{figure*}

The STM tip is an integral part of our atomic-scale Josephson junctions. To assure that the main findings remain valid independent of details of the tip apex, we investigated several tips obtained through reshaping by major tip indentations into the Pb substrate. Supplementary Fig.\ \ref{Fig:Sfig4} shows the non-reciprocity of the retrapping current as a function of \Gpd\ for junctions including Cr and Mn atoms, but measured with different tips. All tips show a positive value of the asymmetry $\Delta I_\mathrm{re}=|I_\mathrm{re,+}| -| I_\mathrm{re,-}|$ in case of Cr, a negative value for Mn, and no asymmetry for Pb adatoms. While these qualitative observations are robust for all tips, there are small differences in the magnitude of the asymmetry at the same value of \Gpd. We tentatively ascribe these variations to tip-dependent Josephson coupling energies and quasiparticle currents as well as noise levels.

\section{Supplementary Note 5: Theoretical simulations}

\begin{figure*}[htp]\centering
\includegraphics[width=0.95\linewidth]{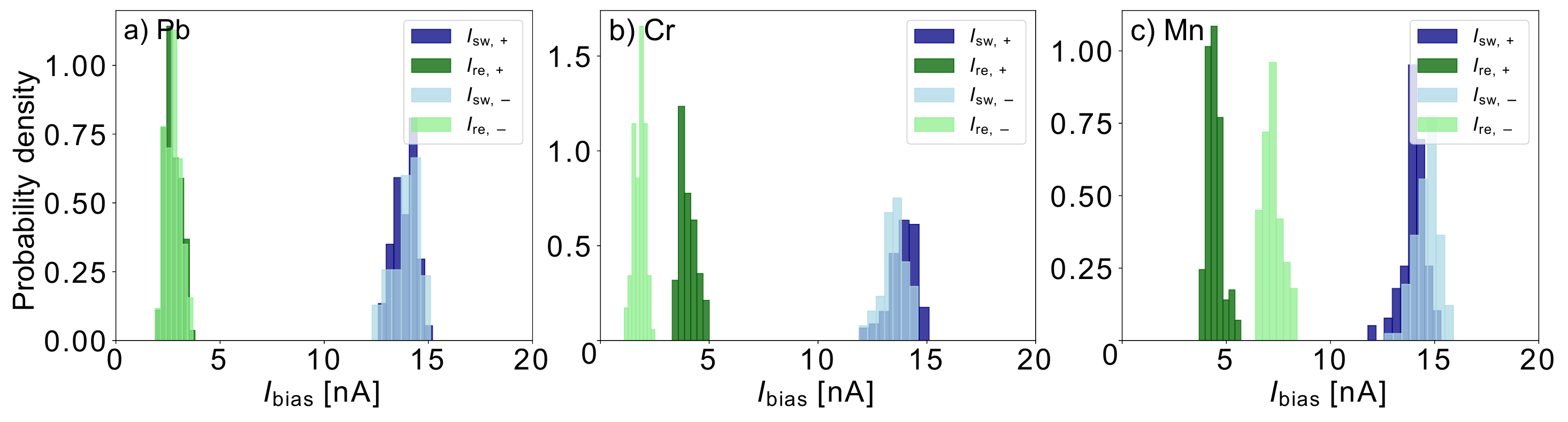}
\caption{\textbf{Statistics of switching and retrapping currents (theoretical simulations based on $I-V$ measurements). } (a-c) Histograms of absolute values of switching and retrapping currents for the two bias directions, as extracted from individual $V-I$ curves from simulation of Eq. \eqref{eq:eom_KM} with $I_\textrm{qp}(V)$ obtained from experimental $I-V$ curves of a (a) Pb, (b) Cr, and (c) Mn junction at $G_\mathrm{N} = 50$ $\mu$S (c.f. Fig.\ 3f of the main text as well as Eq.\ (\ref{eq:fit}) and corresponding text). Each histogram includes data extracted from 100 sweeps for each current direction. For parameters, see text below Eq. \eqref{eq:eom_KM}. 
}
\label{Fig:Sfig5}
\end{figure*}

Our theoretical simulations underlying Fig.\ 4 of the main text are based on the RCSJ model for a current biased junction \cite{SAmbegaokar1969,SIvanchenko1969},
\begin{equation}
    I_\textrm{bias} = C \frac{\mathrm{d}}{\mathrm{d}t} V + I_\mathrm{s}(\varphi) + I_\mathrm{d}(V) + \delta I.
    \label{SEq:RSCJ}
\end{equation}
Here, $I_\textrm{bias}$ is the current bias, $V$ the voltage drop at the junction, $C$ the junction capacitance, and $\varphi$ the phase difference across the junction. We assume a symmetric and sinusoidal current-phase relation $I_\mathrm{s}(\varphi) = I_c \sin\varphi$. We allow for a general nonlinear dissipative current $I_\mathrm{d}(V)$, with associated Nyquist noise $\delta I$ with correlator $\ev{\delta I(t) \delta I(t')} \propto \delta(t-t')$ (see below). When combined with the Josephson relation $V = (\hbar/2e) \mathrm{d}\varphi/\mathrm{d}t$, Eq.\ (\ref{SEq:RSCJ}) gives a Langevin equation for the phase difference across the junction. We solve the Langevin equation by Monte Carlo integration, accounting for the current sweep, to obtain the results shown in Fig. 4 of the main text as well as in the Supplementary Figs.\ \ref{Fig:Sfig5} and \ref{Fig:Sfig6}. 

The dissipative current $I_\mathrm{d}(V)$ includes the quasiparticle current $I_{\textrm{qp}}(V)$ which we extract from experimental $I-V$ traces, see below for details. To account for the observed phase diffusion in the trapped state, we also incorporate frequency-dependent friction. Following Kautz and Martinis \cite{SKautz1990}, we shunt the junction by an additional $RC$ element with Ohmic resistor $\tilde{R}$ and capacitor $\tilde{C}$ to model dissipation induced by the electromagnetic environment. The total dissipative current is then the sum of the quasiparticle current and the current flowing via the $RC$ element,
\begin{equation}
I_\mathrm{d}(V) = I_\textrm{qp}(V) + \frac{V - \tilde{V}}{\tilde{R}},\ \delta I = \delta I_\mathrm{qp} + \delta I_{\tilde{R}}, 
\end{equation}
where $\tilde{V}$ is the voltage drop across the capacitor which satisfies the equation
\begin{equation}
	\frac{\mathrm{d}}{\mathrm{d}t}\tilde{V} = \frac{1}{\tilde{R}\tilde{C}}\left(V - \tilde{V} + \tilde{R}\ \delta I_{\tilde{R}} \right). 
\end{equation}
The $RC$ element is inconsequential at low frequencies (running state), so that damping is dominated by the quasiparticle current. In contrast, it dominates friction at high frequencies (trapped state), allowing for phase diffusion. We assume $V/\tilde{R} \gg I_{\textrm{qp}}(V)$, so that the quasiparticle current is effectively shorted at high frequencies, $I_\mathrm{d}(V) \simeq V/\tilde{R}$. The Nyquist noise associated with the quasiparticle current has correlator $\ev{\delta I_{\textrm{qp}}(t) \delta I_{\textrm{qp}}(t')} =2T [I_{\textrm{qp}}(V)/V] \delta(t-t')$, while the Nyquist noise associated with the resistor $\tilde{R}$ has correlator $\ev{\delta I_{\tilde{R}}(t) \delta I_{\tilde{R}}(t')} =2T \tilde{R}^{-1} \delta(t-t')$. 

Measuring time in units of the inverse plasma frequency, $\tau = \omega_p t$ with $\omega_p = [2eI_c/\hbar C]^{1/2}$, and currents in units of the critical current, $i = I/I_c$, the resulting RCSJ equations become
\begin{equation}\label{eq:eom_KM}
    \frac{\mathrm{d}}{\mathrm{d}\tau}\varphi = v,\ \frac{\mathrm{d}}{\mathrm{d}\tau} v  = i_\mathrm{b} - i_\mathrm{s}(\varphi)  - \left[  i_{\textrm{qp}}\pqty{v} + \frac{v - \tilde{v}}{\tilde{Q}} \right]  - \sqrt{ 2\theta[i_{\textrm{qp}}(v)/v]}\xi_1
    - \sqrt{ 2\tilde{\theta}/\tilde{Q}}\xi_2   ,\ \frac{\mathrm{d}}{\mathrm{d}\tau}\tilde{v} = \frac{1}{\tilde{\tau}} \left(v - \tilde{v} + \sqrt{2\tilde{\theta} \tilde{Q}}\xi_2\right),
\end{equation}
where we defined the dimensionless voltages $v = 2eV/\hbar\omega_p$ and $\tilde{v} = 2e \tilde{V} /\hbar \omega_p$, the dimensionless currents $i_\mathrm{b} = I_\mathrm{bias} /I_c$, $i_\mathrm{s}(\varphi) = I_\mathrm{s}(\varphi)/I_c = \sin \varphi$ and  $i_{\textrm{qp}}(v) = I_{\textrm{qp}}(\hbar\omega_p v/2e)/I_c$, the effective quality factor $\tilde{Q} = \tilde{R}C\omega_p $ at large frequencies, as well as the reduced temperatures ${\theta} = {T}/E_J$ and $\tilde{\theta} = \tilde{T}/E_J$. (Here, $E_J=\hbar I_c/2e$ is the Josephson energy and $\tilde{T}$ is the temperature of the resistor $\tilde{R}$.) We also defined dimensionless Langevin currents $\xi_1$ and $\xi_2$ with normalized correlations $\ev{\xi_i(\tau) \xi_j(\tau')} = \delta_{ij}\delta(\tau-\tau')$ corresponding to $\delta I_\mathrm{qp}$ and $\delta I_{\tilde{R}}$, respectively. 
We estimate the experimental parameters as $R_{\textrm{N}} \sim 20$ k$\Omega$, $\Delta \sim 1.5$ meV, $T \sim 0.1$ meV and $C \sim 10^{-15}$ F. This gives $I_c \sim 100$ nA, $E_J \sim 0.2$ meV and $\hbar\omega_p \sim  0.3$ meV. The reduced temperature is thus $\theta = 0.5$. For the RC element we choose parameters $\tilde Q=10$, $\tilde\tau=1000$, and $\tilde\theta=\theta$. We sweep the bias current with a rate $\mathrm{d} I_\textrm{bias}/\mathrm{d}t = 10^{-7} I_c \omega_p \sim 1$ nA/$\mu$s. The experimental sweep rate is smaller by about a factor of $10^{-3}$, but this would make the numerical simulations forbidding. Along with the simplified current-phase relation and the order-of-magnitude estimates of experimental parameters, this implies that one can only expect qualitative, but not quantitative agreement between simulations and experiment. 

The results of the theoretical simulations are summarized in Fig.\ 4 of the main text as well as Supplementary Fig.\ \ref{Fig:Sfig5}. Here we discuss the latter, which shows histograms of the absolute values of switching and retrapping currents extracted from 100 sweeps in each current direction. Note that the panels only differ in the precise form of $I_{\textrm{qp}} (V)$ which is extracted from the $I-V$ curves of Pb, Cr and Mn, respectively.  The simulations based on the $I_\textrm{qp}(V)$ of Pb do not show  asymmetry in the switching or the retrapping currents. The simulations based on the $I_\textrm{qp}(V)$ of Cr and Mn exhibit weak asymmetry in the switching currents and strong asymmetry in the retrapping currents, correctly reproducing the qualitative features of the experimental histograms in Supplementary Figs.\ 2-4. 

\subsection{Asymmetric current-phase relation}

\begin{figure*}[htp]\centering
\includegraphics[width=0.35\linewidth]{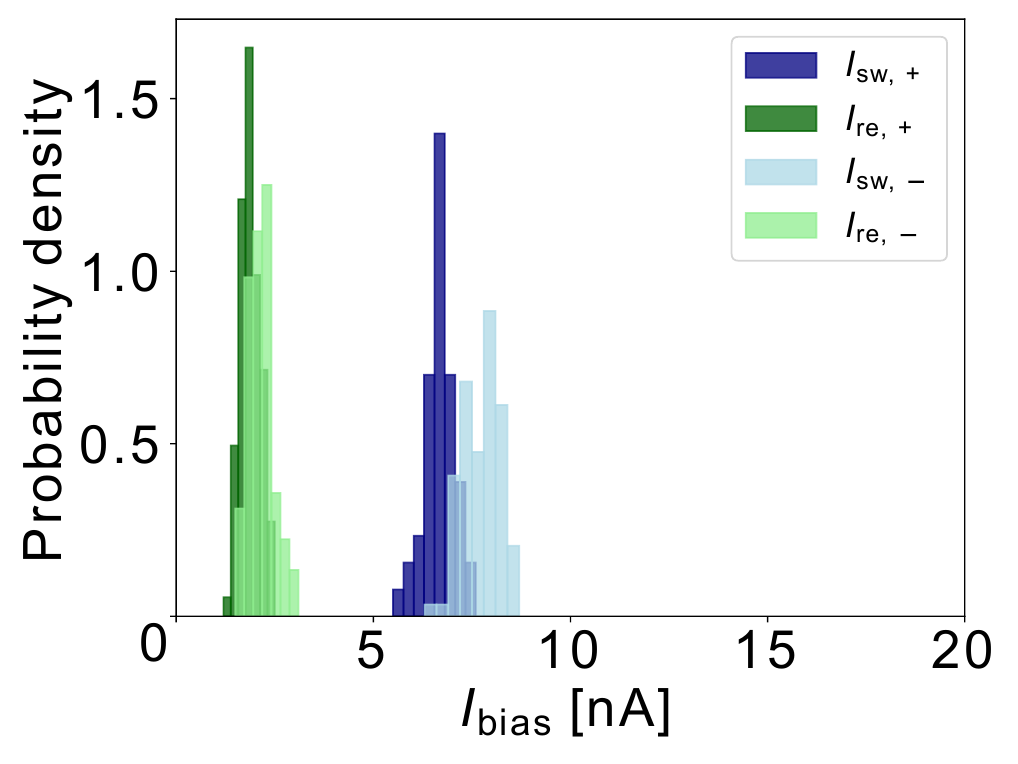}
\caption{\textbf{Simulated statistics of switching and retrapping currents with symmetric quasiparticle current and asymmetric current-phase relation.} Histograms of absolute values of switching and retrapping currents for the two bias directions, as extracted from individual $V-I$ curves in simulations of Eq.\ \eqref{eq:eom_KM}. $I_\textrm{qp}(V)$ is obtained from experimental $I-V$ curves of a Pb junction at $G_\mathrm{N} = 50$ $\mu$S. The asymmetric current-phase relation is given in Eq.\ \eqref{eq:asym_curr_phase}. Each histogram includes data extracted from 98 sweeps for each current direction. Other parameters as in Supplementary Fig.\ \ref{Fig:Sfig5}.}
\label{Fig:Sfig6}
\end{figure*}

To rule out the possibility that the observed asymmetry stems from the current-phase relation $I_\mathrm{s}(\varphi)$ rather than from the dissipative quasiparticle current, we now demonstrate that an asymmetric current-phase relation leads to strong asymmetry in the switching currents and weak asymmetry in the retrapping currents, contrasting with our experimental observations. To this end, we simulate Eq.\ \eqref{eq:eom_KM} using Pb $I-V$ data for $I_{\textrm{qp}}(V)$ together with an asymmetric current-phase relation 
\begin{equation}\label{eq:asym_curr_phase}
	I_\mathrm{s}(\varphi) = I_0 \left[ \sin(\varphi - \varphi_0) + b \sin(2\varphi) \right]. 
\end{equation}
We choose $\varphi_0 = 0.5 = b$ and fix $I_0\simeq 54.2$ nA by requiring that the current entering the definition of the plasma frequency, i.e., the slope of $I_\mathrm{s}$ around the stable minimum, is still $100$ nA (which we continue to use as the unit of current). The critical current now depends on direction, with $I_{c,+} \simeq 53.3$ nA and $I_{c,-} \simeq 80.0$ nA. Histograms of switching and retrapping currents obtained by simulating  Eq.\ \eqref{eq:eom_KM} with the current-phase relation given in Eq.\ \eqref{eq:asym_curr_phase} are presented in Supplementary Fig. \ref{Fig:Sfig6}. The asymmetry of the switching currents is clearly much greater than that of the retrapping currents. Thus, a symmetric dissipative current together with an asymmetric current-phase relation cannot explain the phenomenology of strongly asymmetric retrapping currents and weakly asymmetric switching currents observed for the Cr and Mn Josephson junctions. 

\subsection{Extraction of quasiparticle current}\label{sec:extract_qp_current}

\begin{table}[hb]
\begin{tabular}{|c||c|c|c|c|c|c|c|c|}
\hline
 & $V_{\textrm{offset}}$ [mV] & $I_{\textrm{offset}}$ [nA] & $\delta V$ [mV] & $A$ [nA] & $B$ [nA] & $C$ [nA/mV] & $D$ [nA/mV$^2$] & $E$ [nA/mV$^3$]    \\ \hline  
 Pb & 0.0187 & 0.0292 & 0.135 & 16.4 & -20.6 & 7.00 & 0.121 & -21.8 \\ \hline 
 Cr & 0.0210 & 0.00169 & 0.140 & 5.96 & -7.71 & 4.01 & 2.47 & -1.01 \\ \hline 
 Mn & 0.0129 & -0.123 & 0.138 & 8.47 & -10.7 & 3.66 & -2.52 & 15.1 \\ \hline 
\end{tabular}
\caption{\textbf{Quasiparticle current fitting parameters.} Fitting parameters for extracting the quasiparticle current $I_{\textrm{qp}}(V)$ from the measured current $I_{\textrm{meas}}(V)$ by subtracting the Josephson peak due to incoherent Cooper-pair tunneling, see Eq.\ (\ref{eq:fit}) and corresponding text. }
\label{tab:fit}
\end{table}

We extract the quasiparticle contribution to the dissipative current $I_{\textrm{qp}}(V)$ from voltage-biased measurements of Pb, Cr, and Mn junctions at the normal-state conductance of $G_\mathrm{N} = 50$ $\mu$S, see Fig.\ 3f of the main text. In addition to the quasiparticle current, these traces include a Josephson peak originating from incoherent Cooper-pair tunneling. We remove the Josephson contribution $I_{\textrm{J}}(V)$ by fitting to the phenomenological expressions \cite{SGrabert1999}
\begin{subequations}\label{eq:fit}
\begin{align}
    I_{\textrm{meas}}(V) =&\ I_{\textrm{J}}(V + V_{\textrm{offset}}) + I_{\textrm{qp,0}}(V + V_{\textrm{offset}}) + I_{\textrm{offset}},\\
    I_{\textrm{J}}(V) =&\ A \frac{ V \delta V }{V^2 + \delta V^2} + B \frac{V^3 \delta V}{(V^2 + \delta V^2)^2},\\ 
    I_{\textrm{qp,0}}(V) =&\ C V + D V^2  + E V^3,    
\end{align}
\end{subequations}
over a voltage range $e|V| \ll \Delta$, which contains the Josephson peak. (We choose
$e|V| < 0.32$ meV.) We account also for offsets in the measured voltage and current through the parameters $V_{\textrm{offset}}$ and $I_{\textrm{offset}}$. The fit parameters are collected in Table I. We then subtract the Josephson contribution as well as the offsets from the measured data to isolate the quasiparticle contribution. To reduce the fluctuations at small $V$ associated with the Josephson contribution, a Gaussian filter (width $\sigma = 5$ datapoints $\simeq 0.55$ mV) is applied to the isolated quasiparticle current data. Finally, $I_{\textrm{qp}}(V)$ is obtained by interpolation using a linear splining procedure, enforcing $I_\mathrm{qp}(0) = 0$. 

\def\urlprefix{}
 \def\url#1.{}

\begin{thebibliography}{44}%
\makeatletter
\providecommand \@ifxundefined [1]{%
 \@ifx{#1\undefined}
}%
\providecommand \@ifnum [1]{%
 \ifnum #1\expandafter \@firstoftwo
 \else \expandafter \@secondoftwo
 \fi
}%
\providecommand \@ifx [1]{%
 \ifx #1\expandafter \@firstoftwo
 \else \expandafter \@secondoftwo
 \fi
}%
\providecommand \natexlab [1]{#1}%
\providecommand \enquote  [1]{``#1''}%
\providecommand \bibnamefont  [1]{#1}%
\providecommand \bibfnamefont [1]{#1}%
\providecommand \citenamefont [1]{#1}%
\providecommand \href@noop [0]{\@secondoftwo}%
\providecommand \href [0]{\begingroup \@sanitize@url \@href}%
\providecommand \@href[1]{\@@startlink{#1}\@@href}%
\providecommand \@@href[1]{\endgroup#1\@@endlink}%
\providecommand \@sanitize@url [0]{\catcode `\\12\catcode `\$12\catcode
  `\&12\catcode `\#12\catcode `\^12\catcode `\_12\catcode `\%12\relax}%
\providecommand \@@startlink[1]{}%
\providecommand \@@endlink[0]{}%
\providecommand \url  [0]{\begingroup\@sanitize@url \@url }%
\providecommand \@url [1]{\endgroup\@href {#1}{\urlprefix }}%
\providecommand \urlprefix  [0]{URL }%
\providecommand \Eprint [0]{\href }%
\providecommand \doibase [0]{https://doi.org/}%
\providecommand \selectlanguage [0]{\@gobble}%
\providecommand \bibinfo  [0]{\@secondoftwo}%
\providecommand \bibfield  [0]{\@secondoftwo}%
\providecommand \translation [1]{[#1]}%
\providecommand \BibitemOpen [0]{}%
\providecommand \bibitemStop [0]{}%
\providecommand \bibitemNoStop [0]{.\EOS\space}%
\providecommand \EOS [0]{\spacefactor3000\relax}%
\providecommand \BibitemShut  [1]{\csname bibitem#1\endcsname}%
\let\auto@bib@innerbib\@empty
\bibitem [{\citenamefont {Scaff}\ and\ \citenamefont {Ohl}(1947)}]{Scaff1947}%
  \BibitemOpen
  \bibfield  {author} {\bibinfo {author} {\bibfnamefont {J.~H.}\ \bibnamefont
  {Scaff}}\ and\ \bibinfo {author} {\bibfnamefont {R.~S.}\ \bibnamefont
  {Ohl}},\ }\bibfield  {title} {\bibinfo {title} {{Development of Silicon
  Crystal Rectifiers for Microwave Radar Receivers}},\ }\href
  {https://doi.org/https://doi.org/10.1002/j.1538-7305.1947.tb01310.x}
  {\bibfield  {journal} {\bibinfo  {journal} {Bell Syst. Tech. J.}\ }\textbf
  {\bibinfo {volume} {26}},\ \bibinfo {pages} {1} (\bibinfo {year}
  {1947})}\BibitemShut {NoStop}%
\bibitem [{\citenamefont {Shockley}(1949)}]{Shockley1949}%
  \BibitemOpen
  \bibfield  {author} {\bibinfo {author} {\bibfnamefont {W.}~\bibnamefont
  {Shockley}},\ }\bibfield  {title} {\bibinfo {title} {{The Theory of p-n
  Junctions in Semiconductors and p-n Junction Transistors}},\ }\href
  {https://doi.org/https://doi.org/10.1002/j.1538-7305.1949.tb03645.x}
  {\bibfield  {journal} {\bibinfo  {journal} {Bell Syst. Tech. J.}\ }\textbf
  {\bibinfo {volume} {28}},\ \bibinfo {pages} {435} (\bibinfo {year}
  {1949})}\BibitemShut {NoStop}%
\bibitem [{\citenamefont {Tokura}\ and\ \citenamefont
  {Nagaosa}(2018)}]{Tokura2018}%
  \BibitemOpen
  \bibfield  {author} {\bibinfo {author} {\bibfnamefont {Y.}~\bibnamefont
  {Tokura}}\ and\ \bibinfo {author} {\bibfnamefont {N.}~\bibnamefont
  {Nagaosa}},\ }\bibfield  {title} {\bibinfo {title} {Nonreciprocal responses
  from non-centrosymmetric quantum materials},\ }\href
  {https://doi.org/10.1038/s41467-018-05759-4} {\bibfield  {journal} {\bibinfo
  {journal} {Nature Commun.}\ }\textbf {\bibinfo {volume} {9}},\ \bibinfo
  {pages} {1} (\bibinfo {year} {2018})}\BibitemShut {NoStop}%
\bibitem [{\citenamefont {Rikken}\ and\ \citenamefont
  {Wyder}(2005)}]{Rikken2005}%
  \BibitemOpen
  \bibfield  {author} {\bibinfo {author} {\bibfnamefont {G.~L. J.~A.}\
  \bibnamefont {Rikken}}\ and\ \bibinfo {author} {\bibfnamefont
  {P.}~\bibnamefont {Wyder}},\ }\bibfield  {title} {\bibinfo {title}
  {Magnetoelectric anisotropy in diffusive transport},\ }\href
  {https://doi.org/10.1103/PhysRevLett.94.016601} {\bibfield  {journal}
  {\bibinfo  {journal} {Phys. Rev. Lett.}\ }\textbf {\bibinfo {volume} {94}},\
  \bibinfo {pages} {016601} (\bibinfo {year} {2005})}\BibitemShut {NoStop}%
\bibitem [{\citenamefont {Wakatsuki}\ \emph {et~al.}(2017)\citenamefont
  {Wakatsuki}, \citenamefont {Saito}, \citenamefont {Hoshino}, \citenamefont
  {Itahashi}, \citenamefont {Ideue}, \citenamefont {Ezawa}, \citenamefont
  {Iwasa},\ and\ \citenamefont {Nagaosa}}]{Wakatsuki2017}%
  \BibitemOpen
  \bibfield  {author} {\bibinfo {author} {\bibfnamefont {R.}~\bibnamefont
  {Wakatsuki}}, \bibinfo {author} {\bibfnamefont {Y.}~\bibnamefont {Saito}},
  \bibinfo {author} {\bibfnamefont {S.}~\bibnamefont {Hoshino}}, \bibinfo
  {author} {\bibfnamefont {Y.~M.}\ \bibnamefont {Itahashi}}, \bibinfo {author}
  {\bibfnamefont {T.}~\bibnamefont {Ideue}}, \bibinfo {author} {\bibfnamefont
  {M.}~\bibnamefont {Ezawa}}, \bibinfo {author} {\bibfnamefont
  {Y.}~\bibnamefont {Iwasa}},\ and\ \bibinfo {author} {\bibfnamefont
  {N.}~\bibnamefont {Nagaosa}},\ }\bibfield  {title} {\bibinfo {title}
  {Nonreciprocal charge transport in noncentrosymmetric superconductors},\
  }\href {https://doi.org/10.1126/sciadv.1602390} {\bibfield  {journal}
  {\bibinfo  {journal} {Science Adv.}\ }\textbf {\bibinfo {volume} {3}},\
  \bibinfo {pages} {e1602390} (\bibinfo {year} {2017})}\BibitemShut {NoStop}%
\bibitem [{\citenamefont {Qin}\ \emph {et~al.}(2017)\citenamefont {Qin},
  \citenamefont {Shi}, \citenamefont {Ideue}, \citenamefont {Yoshida},
  \citenamefont {Zak}, \citenamefont {Tenne}, \citenamefont {Kikitsu},
  \citenamefont {Inoue}, \citenamefont {Hashizume},\ and\ \citenamefont
  {Iwasa}}]{Qin2017}%
  \BibitemOpen
  \bibfield  {author} {\bibinfo {author} {\bibfnamefont {F.}~\bibnamefont
  {Qin}}, \bibinfo {author} {\bibfnamefont {W.}~\bibnamefont {Shi}}, \bibinfo
  {author} {\bibfnamefont {T.}~\bibnamefont {Ideue}}, \bibinfo {author}
  {\bibfnamefont {M.}~\bibnamefont {Yoshida}}, \bibinfo {author} {\bibfnamefont
  {A.}~\bibnamefont {Zak}}, \bibinfo {author} {\bibfnamefont {R.}~\bibnamefont
  {Tenne}}, \bibinfo {author} {\bibfnamefont {T.}~\bibnamefont {Kikitsu}},
  \bibinfo {author} {\bibfnamefont {D.}~\bibnamefont {Inoue}}, \bibinfo
  {author} {\bibfnamefont {D.}~\bibnamefont {Hashizume}},\ and\ \bibinfo
  {author} {\bibfnamefont {Y.}~\bibnamefont {Iwasa}},\ }\bibfield  {title}
  {\bibinfo {title} {Superconductivity in a chiral nanotube},\ }\href@noop {}
  {\bibfield  {journal} {\bibinfo  {journal} {Nature Commun.}\ }\textbf
  {\bibinfo {volume} {8}},\ \bibinfo {pages} {1} (\bibinfo {year}
  {2017})}\BibitemShut {NoStop}%
\bibitem [{\citenamefont {Bauriedl}\ \emph {et~al.}(2022)\citenamefont
  {Bauriedl}, \citenamefont {Bäuml}, \citenamefont {Fuchs}, \citenamefont
  {Baumgartner}, \citenamefont {Paulik}, \citenamefont {Bauer}, \citenamefont
  {Lin}, \citenamefont {Lupton}, \citenamefont {Taniguchi}, \citenamefont
  {Watanabe}, \citenamefont {Strunk},\ and\ \citenamefont
  {Paradiso}}]{Bauriedl2022}%
  \BibitemOpen
  \bibfield  {author} {\bibinfo {author} {\bibfnamefont {L.}~\bibnamefont
  {Bauriedl}}, \bibinfo {author} {\bibfnamefont {C.}~\bibnamefont {Bäuml}},
  \bibinfo {author} {\bibfnamefont {L.}~\bibnamefont {Fuchs}}, \bibinfo
  {author} {\bibfnamefont {C.}~\bibnamefont {Baumgartner}}, \bibinfo {author}
  {\bibfnamefont {N.}~\bibnamefont {Paulik}}, \bibinfo {author} {\bibfnamefont
  {J.~M.}\ \bibnamefont {Bauer}}, \bibinfo {author} {\bibfnamefont {K.-Q.}\
  \bibnamefont {Lin}}, \bibinfo {author} {\bibfnamefont {J.~M.}\ \bibnamefont
  {Lupton}}, \bibinfo {author} {\bibfnamefont {T.}~\bibnamefont {Taniguchi}},
  \bibinfo {author} {\bibfnamefont {K.}~\bibnamefont {Watanabe}}, \bibinfo
  {author} {\bibfnamefont {C.}~\bibnamefont {Strunk}},\ and\ \bibinfo {author}
  {\bibfnamefont {N.}~\bibnamefont {Paradiso}},\ }\bibfield  {title} {\bibinfo
  {title} {{Supercurrent diode effect and magnetochiral anisotropy in few-layer
  {NbSe}$_2$}},\ }\href {https://doi.org/10.1038%2Fs41467-022-31954-5}
  {\bibfield  {journal} {\bibinfo  {journal} {Nature Commun.}\ }\textbf
  {\bibinfo {volume} {13}},\ \bibinfo {pages} {4266} (\bibinfo {year}
  {2022})}\BibitemShut {NoStop}%
\bibitem [{\citenamefont {Ando}\ \emph {et~al.}(2020)\citenamefont {Ando},
  \citenamefont {Miyasaka}, \citenamefont {Li}, \citenamefont {Ishizuka},
  \citenamefont {Arakawa}, \citenamefont {Shiota}, \citenamefont {Moriyama},
  \citenamefont {Yanase},\ and\ \citenamefont {Ono}}]{Ando2020}%
  \BibitemOpen
  \bibfield  {author} {\bibinfo {author} {\bibfnamefont {F.}~\bibnamefont
  {Ando}}, \bibinfo {author} {\bibfnamefont {Y.}~\bibnamefont {Miyasaka}},
  \bibinfo {author} {\bibfnamefont {T.}~\bibnamefont {Li}}, \bibinfo {author}
  {\bibfnamefont {J.}~\bibnamefont {Ishizuka}}, \bibinfo {author}
  {\bibfnamefont {T.}~\bibnamefont {Arakawa}}, \bibinfo {author} {\bibfnamefont
  {Y.}~\bibnamefont {Shiota}}, \bibinfo {author} {\bibfnamefont
  {T.}~\bibnamefont {Moriyama}}, \bibinfo {author} {\bibfnamefont
  {Y.}~\bibnamefont {Yanase}},\ and\ \bibinfo {author} {\bibfnamefont
  {T.}~\bibnamefont {Ono}},\ }\bibfield  {title} {\bibinfo {title} {Observation
  of superconducting diode effect},\ }\href
  {https://doi.org/10.1038/s41586-020-2590-4} {\bibfield  {journal} {\bibinfo
  {journal} {Nature}\ }\textbf {\bibinfo {volume} {584}},\ \bibinfo {pages}
  {373} (\bibinfo {year} {2020})}\BibitemShut {NoStop}%
\bibitem [{\citenamefont {Narita}\ \emph {et~al.}(2022)\citenamefont {Narita},
  \citenamefont {Ishizuka}, \citenamefont {Kawarazaki}, \citenamefont {Kan},
  \citenamefont {Shiota}, \citenamefont {Moriyama}, \citenamefont {Shimakawa},
  \citenamefont {Ognev}, \citenamefont {Samardak}, \citenamefont {Yanase},\
  and\ \citenamefont {Ono}}]{Narita2022}%
  \BibitemOpen
  \bibfield  {author} {\bibinfo {author} {\bibfnamefont {H.}~\bibnamefont
  {Narita}}, \bibinfo {author} {\bibfnamefont {J.}~\bibnamefont {Ishizuka}},
  \bibinfo {author} {\bibfnamefont {R.}~\bibnamefont {Kawarazaki}}, \bibinfo
  {author} {\bibfnamefont {D.}~\bibnamefont {Kan}}, \bibinfo {author}
  {\bibfnamefont {Y.}~\bibnamefont {Shiota}}, \bibinfo {author} {\bibfnamefont
  {T.}~\bibnamefont {Moriyama}}, \bibinfo {author} {\bibfnamefont
  {Y.}~\bibnamefont {Shimakawa}}, \bibinfo {author} {\bibfnamefont {A.~V.}\
  \bibnamefont {Ognev}}, \bibinfo {author} {\bibfnamefont {A.~S.}\ \bibnamefont
  {Samardak}}, \bibinfo {author} {\bibfnamefont {Y.}~\bibnamefont {Yanase}},\
  and\ \bibinfo {author} {\bibfnamefont {T.}~\bibnamefont {Ono}},\ }\bibfield
  {title} {\bibinfo {title} {Field-free superconducting diode effect in
  noncentrosymmetric superconductor/ferromagnet multilayers},\ }\href
  {https://doi.org/10.1038/s41565-022-01159-4} {\bibfield  {journal} {\bibinfo
  {journal} {Nat. Nanotechnol.}\ }\textbf {\bibinfo {volume} {17}},\ \bibinfo
  {pages} {823–828} (\bibinfo {year} {2022})}\BibitemShut {NoStop}%
\bibitem [{\citenamefont {Baumgartner}\ \emph {et~al.}(2021)\citenamefont
  {Baumgartner}, \citenamefont {Fuchs}, \citenamefont {Costa}, \citenamefont
  {Reinhardt}, \citenamefont {Gronin}, \citenamefont {Gardner}, \citenamefont
  {Lindemann}, \citenamefont {Manfra}, \citenamefont {Junior}, \citenamefont
  {Kochan}, \citenamefont {Fabian}, \citenamefont {Paradiso},\ and\
  \citenamefont {Strunk}}]{Baumgartner2021}%
  \BibitemOpen
  \bibfield  {author} {\bibinfo {author} {\bibfnamefont {C.}~\bibnamefont
  {Baumgartner}}, \bibinfo {author} {\bibfnamefont {L.}~\bibnamefont {Fuchs}},
  \bibinfo {author} {\bibfnamefont {A.}~\bibnamefont {Costa}}, \bibinfo
  {author} {\bibfnamefont {S.}~\bibnamefont {Reinhardt}}, \bibinfo {author}
  {\bibfnamefont {S.}~\bibnamefont {Gronin}}, \bibinfo {author} {\bibfnamefont
  {G.~C.}\ \bibnamefont {Gardner}}, \bibinfo {author} {\bibfnamefont
  {T.}~\bibnamefont {Lindemann}}, \bibinfo {author} {\bibfnamefont {M.~J.}\
  \bibnamefont {Manfra}}, \bibinfo {author} {\bibfnamefont {P.~E.~F.}\
  \bibnamefont {Junior}}, \bibinfo {author} {\bibfnamefont {D.}~\bibnamefont
  {Kochan}}, \bibinfo {author} {\bibfnamefont {J.}~\bibnamefont {Fabian}},
  \bibinfo {author} {\bibfnamefont {N.}~\bibnamefont {Paradiso}},\ and\
  \bibinfo {author} {\bibfnamefont {C.}~\bibnamefont {Strunk}},\ }\bibfield
  {title} {\bibinfo {title} {{Supercurrent rectification and magnetochiral
  effects in symmetric Josephson junctions}},\ }\href
  {https://doi.org/10.1038/s41565-021-01009-9} {\bibfield  {journal} {\bibinfo
  {journal} {Nat. Nanotechnol.}\ }\textbf {\bibinfo {volume} {17}},\ \bibinfo
  {pages} {39} (\bibinfo {year} {2021})}\BibitemShut {NoStop}%
\bibitem [{\citenamefont {Pal}\ \emph {et~al.}()\citenamefont {Pal},
  \citenamefont {Chakraborty}, \citenamefont {Sivakumar}, \citenamefont
  {Davydova}, \citenamefont {Gopi}, \citenamefont {Pandeya}, \citenamefont
  {Krieger}, \citenamefont {Zhang}, \citenamefont {Date}, \citenamefont {Ju},
  \citenamefont {Yuan}, \citenamefont {Schröter}, \citenamefont {Fu},\ and\
  \citenamefont {Parkin}}]{Pal2021}%
  \BibitemOpen
  \bibfield  {author} {\bibinfo {author} {\bibfnamefont {B.}~\bibnamefont
  {Pal}}, \bibinfo {author} {\bibfnamefont {A.}~\bibnamefont {Chakraborty}},
  \bibinfo {author} {\bibfnamefont {P.~K.}\ \bibnamefont {Sivakumar}}, \bibinfo
  {author} {\bibfnamefont {M.}~\bibnamefont {Davydova}}, \bibinfo {author}
  {\bibfnamefont {A.~K.}\ \bibnamefont {Gopi}}, \bibinfo {author}
  {\bibfnamefont {A.~K.}\ \bibnamefont {Pandeya}}, \bibinfo {author}
  {\bibfnamefont {J.~A.}\ \bibnamefont {Krieger}}, \bibinfo {author}
  {\bibfnamefont {Y.}~\bibnamefont {Zhang}}, \bibinfo {author} {\bibfnamefont
  {M.}~\bibnamefont {Date}}, \bibinfo {author} {\bibfnamefont {S.}~\bibnamefont
  {Ju}}, \bibinfo {author} {\bibfnamefont {N.}~\bibnamefont {Yuan}}, \bibinfo
  {author} {\bibfnamefont {N.~B.~M.}\ \bibnamefont {Schröter}}, \bibinfo
  {author} {\bibfnamefont {L.}~\bibnamefont {Fu}},\ and\ \bibinfo {author}
  {\bibfnamefont {S.~S.~P.}\ \bibnamefont {Parkin}},\ }\href
  {https://doi.org/10.48550/ARXIV.2112.11285} {\bibinfo {title} {{Josephson
  diode effect from Cooper pair momentum in a topological semimetal}}},\
  \bibinfo {howpublished} {arXiv:2112.11285 (2021)}\BibitemShut {NoStop}%
\bibitem [{\citenamefont {Diez-Merida}\ \emph {et~al.}()\citenamefont
  {Diez-Merida}, \citenamefont {Diez-Carlon}, \citenamefont {Yang},
  \citenamefont {Xie}, \citenamefont {Gao}, \citenamefont {Watanabe},
  \citenamefont {Taniguchi}, \citenamefont {Lu}, \citenamefont {Law},\ and\
  \citenamefont {Efetov}}]{Diez2021}%
  \BibitemOpen
  \bibfield  {author} {\bibinfo {author} {\bibfnamefont {J.}~\bibnamefont
  {Diez-Merida}}, \bibinfo {author} {\bibfnamefont {A.}~\bibnamefont
  {Diez-Carlon}}, \bibinfo {author} {\bibfnamefont {S.~Y.}\ \bibnamefont
  {Yang}}, \bibinfo {author} {\bibfnamefont {Y.~M.}\ \bibnamefont {Xie}},
  \bibinfo {author} {\bibfnamefont {X.~J.}\ \bibnamefont {Gao}}, \bibinfo
  {author} {\bibfnamefont {K.}~\bibnamefont {Watanabe}}, \bibinfo {author}
  {\bibfnamefont {T.}~\bibnamefont {Taniguchi}}, \bibinfo {author}
  {\bibfnamefont {X.}~\bibnamefont {Lu}}, \bibinfo {author} {\bibfnamefont
  {K.~T.}\ \bibnamefont {Law}},\ and\ \bibinfo {author} {\bibfnamefont {D.~K.}\
  \bibnamefont {Efetov}},\ }\href {https://doi.org/10.48550/ARXIV.2110.01067}
  {\bibinfo {title} {{Magnetic Josephson Junctions and Superconducting Diodes
  in Magic Angle Twisted Bilayer Graphene}}},\ \bibinfo {howpublished}
  {arXiv:2110.01067 (2021)}\BibitemShut {NoStop}%
\bibitem [{\citenamefont {Wu}\ \emph {et~al.}(2022)\citenamefont {Wu},
  \citenamefont {Wang}, \citenamefont {Xu}, \citenamefont {Sivakumar},
  \citenamefont {Pasco}, \citenamefont {Filippozzi}, \citenamefont {Parkin},
  \citenamefont {Zeng}, \citenamefont {McQueen},\ and\ \citenamefont
  {Ali}}]{Wu2022}%
  \BibitemOpen
  \bibfield  {author} {\bibinfo {author} {\bibfnamefont {H.}~\bibnamefont
  {Wu}}, \bibinfo {author} {\bibfnamefont {Y.}~\bibnamefont {Wang}}, \bibinfo
  {author} {\bibfnamefont {Y.}~\bibnamefont {Xu}}, \bibinfo {author}
  {\bibfnamefont {P.~K.}\ \bibnamefont {Sivakumar}}, \bibinfo {author}
  {\bibfnamefont {C.}~\bibnamefont {Pasco}}, \bibinfo {author} {\bibfnamefont
  {U.}~\bibnamefont {Filippozzi}}, \bibinfo {author} {\bibfnamefont {S.~S.~P.}\
  \bibnamefont {Parkin}}, \bibinfo {author} {\bibfnamefont {Y.-J.}\
  \bibnamefont {Zeng}}, \bibinfo {author} {\bibfnamefont {T.}~\bibnamefont
  {McQueen}},\ and\ \bibinfo {author} {\bibfnamefont {M.~N.}\ \bibnamefont
  {Ali}},\ }\bibfield  {title} {\bibinfo {title} {{The field-free Josephson
  diode in a van der Waals heterostructure}},\ }\href
  {https://doi.org/10.1038/s41586-022-04504-8} {\bibfield  {journal} {\bibinfo
  {journal} {Nature}\ }\textbf {\bibinfo {volume} {604}},\ \bibinfo {pages}
  {653} (\bibinfo {year} {2022})}\BibitemShut {NoStop}%
\bibitem [{\citenamefont {Bocquillon}\ \emph {et~al.}(2017)\citenamefont
  {Bocquillon}, \citenamefont {Deacon}, \citenamefont {Wiedenmann},
  \citenamefont {Leubner}, \citenamefont {Klapwijk}, \citenamefont {Brüne},
  \citenamefont {Ishibashi}, \citenamefont {Buhmann},\ and\ \citenamefont
  {Molenkamp}}]{Bocquillon2017}%
  \BibitemOpen
  \bibfield  {author} {\bibinfo {author} {\bibfnamefont {E.}~\bibnamefont
  {Bocquillon}}, \bibinfo {author} {\bibfnamefont {R.~S.}\ \bibnamefont
  {Deacon}}, \bibinfo {author} {\bibfnamefont {J.}~\bibnamefont {Wiedenmann}},
  \bibinfo {author} {\bibfnamefont {P.}~\bibnamefont {Leubner}}, \bibinfo
  {author} {\bibfnamefont {T.~M.}\ \bibnamefont {Klapwijk}}, \bibinfo {author}
  {\bibfnamefont {C.}~\bibnamefont {Brüne}}, \bibinfo {author} {\bibfnamefont
  {K.}~\bibnamefont {Ishibashi}}, \bibinfo {author} {\bibfnamefont
  {H.}~\bibnamefont {Buhmann}},\ and\ \bibinfo {author} {\bibfnamefont {L.~W.}\
  \bibnamefont {Molenkamp}},\ }\bibfield  {title} {\bibinfo {title} {{Gapless
  Andreev bound states in the quantum spin Hall insulator HgTe}},\ }\href
  {https://doi.org/10.1038/nnano.2016.159} {\bibfield  {journal} {\bibinfo
  {journal} {Nat. Nanotech.}\ }\textbf {\bibinfo {volume} {12}},\ \bibinfo
  {pages} {137} (\bibinfo {year} {2017})}\BibitemShut {NoStop}%
\bibitem [{\citenamefont {Naaman}\ \emph {et~al.}(2001)\citenamefont {Naaman},
  \citenamefont {Teizer},\ and\ \citenamefont {Dynes}}]{Naaman2001}%
  \BibitemOpen
  \bibfield  {author} {\bibinfo {author} {\bibfnamefont {O.}~\bibnamefont
  {Naaman}}, \bibinfo {author} {\bibfnamefont {W.}~\bibnamefont {Teizer}},\
  and\ \bibinfo {author} {\bibfnamefont {R.~C.}\ \bibnamefont {Dynes}},\
  }\bibfield  {title} {\bibinfo {title} {{Fluctuation Dominated Josephson
  Tunneling with a Scanning Tunneling Microscope}},\ }\href
  {https://doi.org/10.1103/PhysRevLett.87.097004} {\bibfield  {journal}
  {\bibinfo  {journal} {Phys. Rev. Lett.}\ }\textbf {\bibinfo {volume} {87}},\
  \bibinfo {pages} {097004} (\bibinfo {year} {2001})}\BibitemShut {NoStop}%
\bibitem [{\citenamefont {Rodrigo}\ \emph {et~al.}(2004)\citenamefont
  {Rodrigo}, \citenamefont {Suderow},\ and\ \citenamefont
  {Vieira}}]{Rodrigo2004}%
  \BibitemOpen
  \bibfield  {author} {\bibinfo {author} {\bibfnamefont {J.~G.}\ \bibnamefont
  {Rodrigo}}, \bibinfo {author} {\bibfnamefont {H.}~\bibnamefont {Suderow}},\
  and\ \bibinfo {author} {\bibfnamefont {S.}~\bibnamefont {Vieira}},\
  }\bibfield  {title} {\bibinfo {title} {{On the use of STM superconducting
  tips at very low temperatures}},\ }\href
  {https://doi.org/10.1140/epjb/e2004-00273-y} {\bibfield  {journal} {\bibinfo
  {journal} {Eur. Phys. J. B}\ }\textbf {\bibinfo {volume} {40}},\ \bibinfo
  {pages} {483} (\bibinfo {year} {2004})}\BibitemShut {NoStop}%
\bibitem [{\citenamefont {Bastiaans}\ \emph {et~al.}(2019)\citenamefont
  {Bastiaans}, \citenamefont {Cho}, \citenamefont {Chatzopoulos}, \citenamefont
  {Leeuwenhoek}, \citenamefont {Koks},\ and\ \citenamefont
  {Allan}}]{Bastiaans2019}%
  \BibitemOpen
  \bibfield  {author} {\bibinfo {author} {\bibfnamefont {K.~M.}\ \bibnamefont
  {Bastiaans}}, \bibinfo {author} {\bibfnamefont {D.}~\bibnamefont {Cho}},
  \bibinfo {author} {\bibfnamefont {D.}~\bibnamefont {Chatzopoulos}}, \bibinfo
  {author} {\bibfnamefont {M.}~\bibnamefont {Leeuwenhoek}}, \bibinfo {author}
  {\bibfnamefont {C.}~\bibnamefont {Koks}},\ and\ \bibinfo {author}
  {\bibfnamefont {M.~P.}\ \bibnamefont {Allan}},\ }\bibfield  {title} {\bibinfo
  {title} {{Imaging doubled shot noise in a Josephson scanning tunneling
  microscope}},\ }\href {https://doi.org/10.1103/PhysRevB.100.104506}
  {\bibfield  {journal} {\bibinfo  {journal} {Phys. Rev. B}\ }\textbf {\bibinfo
  {volume} {100}},\ \bibinfo {pages} {104506} (\bibinfo {year}
  {2019})}\BibitemShut {NoStop}%
\bibitem [{\citenamefont {Hamidian}\ \emph {et~al.}(2016)\citenamefont
  {Hamidian}, \citenamefont {Edkins}, \citenamefont {Joo}, \citenamefont
  {Kostin}, \citenamefont {Eisaki}, \citenamefont {Uchida}, \citenamefont
  {Lawler}, \citenamefont {Kim}, \citenamefont {Mackenzie}, \citenamefont
  {Fujita}, \citenamefont {Lee},\ and\ \citenamefont {Davis}}]{Hamidian2016}%
  \BibitemOpen
  \bibfield  {author} {\bibinfo {author} {\bibfnamefont {M.~H.}\ \bibnamefont
  {Hamidian}}, \bibinfo {author} {\bibfnamefont {S.~D.}\ \bibnamefont
  {Edkins}}, \bibinfo {author} {\bibfnamefont {S.~H.}\ \bibnamefont {Joo}},
  \bibinfo {author} {\bibfnamefont {A.}~\bibnamefont {Kostin}}, \bibinfo
  {author} {\bibfnamefont {H.}~\bibnamefont {Eisaki}}, \bibinfo {author}
  {\bibfnamefont {S.}~\bibnamefont {Uchida}}, \bibinfo {author} {\bibfnamefont
  {M.~J.}\ \bibnamefont {Lawler}}, \bibinfo {author} {\bibfnamefont {E.-A.}\
  \bibnamefont {Kim}}, \bibinfo {author} {\bibfnamefont {A.~P.}\ \bibnamefont
  {Mackenzie}}, \bibinfo {author} {\bibfnamefont {K.}~\bibnamefont {Fujita}},
  \bibinfo {author} {\bibfnamefont {J.}~\bibnamefont {Lee}},\ and\ \bibinfo
  {author} {\bibfnamefont {J.~C.~S.}\ \bibnamefont {Davis}},\ }\bibfield
  {title} {\bibinfo {title} {{Detection of a Cooper-pair density wave in
  Bi$_2$Sr$_2$CaCu$-2$O$_{8+x}$}},\ }\href
  {https://doi.org/10.1038/nature17411} {\bibfield  {journal} {\bibinfo
  {journal} {Nature}\ }\textbf {\bibinfo {volume} {532}},\ \bibinfo {pages}
  {343} (\bibinfo {year} {2016})}\BibitemShut {NoStop}%
\bibitem [{\citenamefont {Liu}\ \emph {et~al.}(2021)\citenamefont {Liu},
  \citenamefont {Chong}, \citenamefont {Sharma},\ and\ \citenamefont
  {Davis}}]{Liu2021}%
  \BibitemOpen
  \bibfield  {author} {\bibinfo {author} {\bibfnamefont {X.}~\bibnamefont
  {Liu}}, \bibinfo {author} {\bibfnamefont {Y.~X.}\ \bibnamefont {Chong}},
  \bibinfo {author} {\bibfnamefont {R.}~\bibnamefont {Sharma}},\ and\ \bibinfo
  {author} {\bibfnamefont {J.~C.~S.}\ \bibnamefont {Davis}},\ }\bibfield
  {title} {\bibinfo {title} {{Discovery of a Cooper-pair density wave state in
  a transition-metal dichalcogenide}},\ }\href
  {https://doi.org/10.1126/science.abd4607} {\bibfield  {journal} {\bibinfo
  {journal} {Science}\ }\textbf {\bibinfo {volume} {372}},\ \bibinfo {pages}
  {1447} (\bibinfo {year} {2021})}\BibitemShut {NoStop}%
\bibitem [{\citenamefont {J\"ack}\ \emph {et~al.}(2017)\citenamefont {J\"ack},
  \citenamefont {Senkpiel}, \citenamefont {Etzkorn}, \citenamefont {Ankerhold},
  \citenamefont {Ast},\ and\ \citenamefont {Kern}}]{Jaeck2017}%
  \BibitemOpen
  \bibfield  {author} {\bibinfo {author} {\bibfnamefont {B.}~\bibnamefont
  {J\"ack}}, \bibinfo {author} {\bibfnamefont {J.}~\bibnamefont {Senkpiel}},
  \bibinfo {author} {\bibfnamefont {M.}~\bibnamefont {Etzkorn}}, \bibinfo
  {author} {\bibfnamefont {J.}~\bibnamefont {Ankerhold}}, \bibinfo {author}
  {\bibfnamefont {C.~R.}\ \bibnamefont {Ast}},\ and\ \bibinfo {author}
  {\bibfnamefont {K.}~\bibnamefont {Kern}},\ }\bibfield  {title} {\bibinfo
  {title} {{Quantum Brownian Motion at Strong Dissipation Probed by
  Superconducting Tunnel Junctions}},\ }\href
  {https://doi.org/10.1103/PhysRevLett.119.147702} {\bibfield  {journal}
  {\bibinfo  {journal} {Phys. Rev. Lett.}\ }\textbf {\bibinfo {volume} {119}},\
  \bibinfo {pages} {147702} (\bibinfo {year} {2017})}\BibitemShut {NoStop}%
\bibitem [{\citenamefont {Roychowdhury}\ \emph {et~al.}(2015)\citenamefont
  {Roychowdhury}, \citenamefont {Dreyer}, \citenamefont {Anderson},
  \citenamefont {Lobb},\ and\ \citenamefont {Wellstood}}]{Roychowdhury2015}%
  \BibitemOpen
  \bibfield  {author} {\bibinfo {author} {\bibfnamefont {A.}~\bibnamefont
  {Roychowdhury}}, \bibinfo {author} {\bibfnamefont {M.}~\bibnamefont
  {Dreyer}}, \bibinfo {author} {\bibfnamefont {J.~R.}\ \bibnamefont
  {Anderson}}, \bibinfo {author} {\bibfnamefont {C.~J.}\ \bibnamefont {Lobb}},\
  and\ \bibinfo {author} {\bibfnamefont {F.~C.}\ \bibnamefont {Wellstood}},\
  }\bibfield  {title} {\bibinfo {title} {{Microwave Photon-Assisted Incoherent
  Cooper-Pair Tunneling in a Josephson STM}},\ }\href
  {https://doi.org/10.1103/PhysRevApplied.4.034011} {\bibfield  {journal}
  {\bibinfo  {journal} {Phys. Rev. Appl.}\ }\textbf {\bibinfo {volume} {4}},\
  \bibinfo {pages} {034011} (\bibinfo {year} {2015})}\BibitemShut {NoStop}%
\bibitem [{\citenamefont {Kot}\ \emph {et~al.}(2020)\citenamefont {Kot},
  \citenamefont {Drost}, \citenamefont {Uhl}, \citenamefont {Ankerhold},
  \citenamefont {Cuevas},\ and\ \citenamefont {Ast}}]{Kot2020}%
  \BibitemOpen
  \bibfield  {author} {\bibinfo {author} {\bibfnamefont {P.}~\bibnamefont
  {Kot}}, \bibinfo {author} {\bibfnamefont {R.}~\bibnamefont {Drost}}, \bibinfo
  {author} {\bibfnamefont {M.}~\bibnamefont {Uhl}}, \bibinfo {author}
  {\bibfnamefont {J.}~\bibnamefont {Ankerhold}}, \bibinfo {author}
  {\bibfnamefont {J.~C.}\ \bibnamefont {Cuevas}},\ and\ \bibinfo {author}
  {\bibfnamefont {C.~R.}\ \bibnamefont {Ast}},\ }\bibfield  {title} {\bibinfo
  {title} {Microwave-assisted tunneling and interference effects in
  superconducting junctions under fast driving signals},\ }\href
  {https://doi.org/10.1103/PhysRevB.101.134507} {\bibfield  {journal} {\bibinfo
   {journal} {Phys. Rev. B}\ }\textbf {\bibinfo {volume} {101}},\ \bibinfo
  {pages} {134507} (\bibinfo {year} {2020})}\BibitemShut {NoStop}%
\bibitem [{\citenamefont {Peters}\ \emph {et~al.}(2020)\citenamefont {Peters},
  \citenamefont {Bogdanoff}, \citenamefont {Gonz{\'a}lez}, \citenamefont
  {Melischek}, \citenamefont {Simon}, \citenamefont {Reecht}, \citenamefont
  {Winkelmann}, \citenamefont {von Oppen},\ and\ \citenamefont
  {Franke}}]{Peters2020}%
  \BibitemOpen
  \bibfield  {author} {\bibinfo {author} {\bibfnamefont {O.}~\bibnamefont
  {Peters}}, \bibinfo {author} {\bibfnamefont {N.}~\bibnamefont {Bogdanoff}},
  \bibinfo {author} {\bibfnamefont {S.~A.}\ \bibnamefont {Gonz{\'a}lez}},
  \bibinfo {author} {\bibfnamefont {L.}~\bibnamefont {Melischek}}, \bibinfo
  {author} {\bibfnamefont {J.~R.}\ \bibnamefont {Simon}}, \bibinfo {author}
  {\bibfnamefont {G.}~\bibnamefont {Reecht}}, \bibinfo {author} {\bibfnamefont
  {C.~B.}\ \bibnamefont {Winkelmann}}, \bibinfo {author} {\bibfnamefont
  {F.}~\bibnamefont {von Oppen}},\ and\ \bibinfo {author} {\bibfnamefont
  {K.~J.}\ \bibnamefont {Franke}},\ }\bibfield  {title} {\bibinfo {title}
  {{Resonant Andreev reflections probed by photon-assisted tunnelling at the
  atomic scale}},\ }\href
  {https://doi.org/https://doi.org/10.1038/s41567-020-0972-z} {\bibfield
  {journal} {\bibinfo  {journal} {Nature Phys.}\ }\textbf {\bibinfo {volume}
  {16}},\ \bibinfo {pages} {1222} (\bibinfo {year} {2020})}\BibitemShut
  {NoStop}%
\bibitem [{\citenamefont {Randeria}\ \emph {et~al.}(2016)\citenamefont
  {Randeria}, \citenamefont {Feldman}, \citenamefont {Drozdov},\ and\
  \citenamefont {Yazdani}}]{Randeria2016}%
  \BibitemOpen
  \bibfield  {author} {\bibinfo {author} {\bibfnamefont {M.~T.}\ \bibnamefont
  {Randeria}}, \bibinfo {author} {\bibfnamefont {B.~E.}\ \bibnamefont
  {Feldman}}, \bibinfo {author} {\bibfnamefont {I.~K.}\ \bibnamefont
  {Drozdov}},\ and\ \bibinfo {author} {\bibfnamefont {A.}~\bibnamefont
  {Yazdani}},\ }\bibfield  {title} {\bibinfo {title} {{Scanning Josephson
  spectroscopy on the atomic scale}},\ }\href
  {https://doi.org/10.1103/PhysRevB.93.161115} {\bibfield  {journal} {\bibinfo
  {journal} {Phys. Rev. B}\ }\textbf {\bibinfo {volume} {93}},\ \bibinfo
  {pages} {161115} (\bibinfo {year} {2016})}\BibitemShut {NoStop}%
\bibitem [{\citenamefont {Küster}\ \emph {et~al.}(2021)\citenamefont
  {Küster}, \citenamefont {Montero}, \citenamefont {Guimar{\~{a}}es},
  \citenamefont {Brinker}, \citenamefont {Lounis}, \citenamefont {Parkin},\
  and\ \citenamefont {Sessi}}]{Kuester2021}%
  \BibitemOpen
  \bibfield  {author} {\bibinfo {author} {\bibfnamefont {F.}~\bibnamefont
  {Küster}}, \bibinfo {author} {\bibfnamefont {A.~M.}\ \bibnamefont
  {Montero}}, \bibinfo {author} {\bibfnamefont {F.~S.~M.}\ \bibnamefont
  {Guimar{\~{a}}es}}, \bibinfo {author} {\bibfnamefont {S.}~\bibnamefont
  {Brinker}}, \bibinfo {author} {\bibfnamefont {S.}~\bibnamefont {Lounis}},
  \bibinfo {author} {\bibfnamefont {S.~S.~P.}\ \bibnamefont {Parkin}},\ and\
  \bibinfo {author} {\bibfnamefont {P.}~\bibnamefont {Sessi}},\ }\bibfield
  {title} {\bibinfo {title} {{Correlating Josephson supercurrents and Shiba
  states in quantum spins unconventionally coupled to superconductors}},\
  }\href {https://doi.org/10.1038/s41467-021-21347-5} {\bibfield  {journal}
  {\bibinfo  {journal} {Nature Commun.}\ }\textbf {\bibinfo {volume} {12}},\
  \bibinfo {pages} {1108} (\bibinfo {year} {2021})}\BibitemShut {NoStop}%
\bibitem [{\citenamefont {Karan}\ \emph {et~al.}(2022)\citenamefont {Karan},
  \citenamefont {Huang}, \citenamefont {Padurariu}, \citenamefont {Kubala},
  \citenamefont {Theiler}, \citenamefont {Black-Schaffer}, \citenamefont
  {Morr{\'{a}}s}, \citenamefont {Yeyati}, \citenamefont {Cuevas}, \citenamefont
  {Ankerhold}, \citenamefont {Kern},\ and\ \citenamefont {Ast}}]{Karan2022}%
  \BibitemOpen
  \bibfield  {author} {\bibinfo {author} {\bibfnamefont {S.}~\bibnamefont
  {Karan}}, \bibinfo {author} {\bibfnamefont {H.}~\bibnamefont {Huang}},
  \bibinfo {author} {\bibfnamefont {C.}~\bibnamefont {Padurariu}}, \bibinfo
  {author} {\bibfnamefont {B.}~\bibnamefont {Kubala}}, \bibinfo {author}
  {\bibfnamefont {A.}~\bibnamefont {Theiler}}, \bibinfo {author} {\bibfnamefont
  {A.~M.}\ \bibnamefont {Black-Schaffer}}, \bibinfo {author} {\bibfnamefont
  {G.}~\bibnamefont {Morr{\'{a}}s}}, \bibinfo {author} {\bibfnamefont {A.~L.}\
  \bibnamefont {Yeyati}}, \bibinfo {author} {\bibfnamefont {J.~C.}\
  \bibnamefont {Cuevas}}, \bibinfo {author} {\bibfnamefont {J.}~\bibnamefont
  {Ankerhold}}, \bibinfo {author} {\bibfnamefont {K.}~\bibnamefont {Kern}},\
  and\ \bibinfo {author} {\bibfnamefont {C.~R.}\ \bibnamefont {Ast}},\
  }\bibfield  {title} {\bibinfo {title} {Superconducting quantum interference
  at the atomic scale},\ }\href {https://doi.org/10.1038/s41567-022-01644-6}
  {\bibfield  {journal} {\bibinfo  {journal} {Nature Phys.}\ }\textbf {\bibinfo
  {volume} {18}},\ \bibinfo {pages} {893} (\bibinfo {year} {2022})}\BibitemShut
  {NoStop}%
\bibitem [{\citenamefont {Stewart}(1968)}]{Stewart1969}%
  \BibitemOpen
  \bibfield  {author} {\bibinfo {author} {\bibfnamefont {W.~C.}\ \bibnamefont
  {Stewart}},\ }\bibfield  {title} {\bibinfo {title} {{Current-voltage
  characteristics of Josephson junctions}},\ }\href
  {https://doi.org/10.1063/1.1651991} {\bibfield  {journal} {\bibinfo
  {journal} {Appl. Phys. Lett.}\ }\textbf {\bibinfo {volume} {12}},\ \bibinfo
  {pages} {277} (\bibinfo {year} {1968})}\BibitemShut {NoStop}%
\bibitem [{\citenamefont {McCumber}(1968)}]{McCumber1968}%
  \BibitemOpen
  \bibfield  {author} {\bibinfo {author} {\bibfnamefont {D.~E.}\ \bibnamefont
  {McCumber}},\ }\bibfield  {title} {\bibinfo {title} {{Effect of ac Impedance
  on dc Voltage‐Current Characteristics of Superconductor Weak‐Link
  Junctions}},\ }\href {https://doi.org/10.1063/1.1656743} {\bibfield
  {journal} {\bibinfo  {journal} {J. Appl. Phys.}\ }\textbf {\bibinfo {volume}
  {39}},\ \bibinfo {pages} {3113} (\bibinfo {year} {1968})}\BibitemShut
  {NoStop}%
\bibitem [{\citenamefont {Ambegaokar}\ and\ \citenamefont
  {Halperin}(1969)}]{Ambegaokar1969}%
  \BibitemOpen
  \bibfield  {author} {\bibinfo {author} {\bibfnamefont {V.}~\bibnamefont
  {Ambegaokar}}\ and\ \bibinfo {author} {\bibfnamefont {B.~I.}\ \bibnamefont
  {Halperin}},\ }\bibfield  {title} {\bibinfo {title} {{Voltage Due to Thermal
  Noise in the dc Josephson Effect}},\ }\href
  {https://doi.org/10.1103/PhysRevLett.22.1364} {\bibfield  {journal} {\bibinfo
   {journal} {Phys. Rev. Lett.}\ }\textbf {\bibinfo {volume} {22}},\ \bibinfo
  {pages} {1364} (\bibinfo {year} {1969})}\BibitemShut {NoStop}%
\bibitem [{\citenamefont {{Ivanchenko}}\ and\ \citenamefont
  {{Zil'berman}}(1969)}]{Ivanchenko1969}%
  \BibitemOpen
  \bibfield  {author} {\bibinfo {author} {\bibfnamefont {Y.~M.}\ \bibnamefont
  {{Ivanchenko}}}\ and\ \bibinfo {author} {\bibfnamefont {L.~A.}\ \bibnamefont
  {{Zil'berman}}},\ }\bibfield  {title} {\bibinfo {title} {{The Josephson
  Effect in Small Tunnel Contacts}},\ }\href@noop {} {\bibfield  {journal}
  {\bibinfo  {journal} {Sov. JETP}\ }\textbf {\bibinfo {volume} {28}},\
  \bibinfo {pages} {1272} (\bibinfo {year} {1969})}\BibitemShut {NoStop}%
\bibitem [{\citenamefont {Kautz}\ and\ \citenamefont
  {Martinis}(1990)}]{Kautz1990}%
  \BibitemOpen
  \bibfield  {author} {\bibinfo {author} {\bibfnamefont {R.~L.}\ \bibnamefont
  {Kautz}}\ and\ \bibinfo {author} {\bibfnamefont {J.~M.}\ \bibnamefont
  {Martinis}},\ }\bibfield  {title} {\bibinfo {title} {{Noise-affected I-V
  curves in small hysteretic Josephson junctions}},\ }\href
  {https://doi.org/10.1103/PhysRevB.42.9903} {\bibfield  {journal} {\bibinfo
  {journal} {Phys. Rev. B}\ }\textbf {\bibinfo {volume} {42}},\ \bibinfo
  {pages} {9903} (\bibinfo {year} {1990})}\BibitemShut {NoStop}%
\bibitem [{\citenamefont {Ambegaokar}\ and\ \citenamefont
  {Baratoff}(1963)}]{Ambegaokar1963}%
  \BibitemOpen
  \bibfield  {author} {\bibinfo {author} {\bibfnamefont {V.}~\bibnamefont
  {Ambegaokar}}\ and\ \bibinfo {author} {\bibfnamefont {A.}~\bibnamefont
  {Baratoff}},\ }\bibfield  {title} {\bibinfo {title} {{Tunneling Between
  Superconductors}},\ }\href {https://doi.org/10.1103/PhysRevLett.10.486}
  {\bibfield  {journal} {\bibinfo  {journal} {Phys. Rev. Lett.}\ }\textbf
  {\bibinfo {volume} {10}},\ \bibinfo {pages} {486} (\bibinfo {year}
  {1963})}\BibitemShut {NoStop}%
\bibitem [{\citenamefont {Yokoyama}\ \emph {et~al.}(2013)\citenamefont
  {Yokoyama}, \citenamefont {Eto},\ and\ \citenamefont
  {V.~Nazarov}}]{Yokoyama2013}%
  \BibitemOpen
  \bibfield  {author} {\bibinfo {author} {\bibfnamefont {T.}~\bibnamefont
  {Yokoyama}}, \bibinfo {author} {\bibfnamefont {M.}~\bibnamefont {Eto}},\ and\
  \bibinfo {author} {\bibfnamefont {Y.}~\bibnamefont {V.~Nazarov}},\ }\bibfield
   {title} {\bibinfo {title} {{Josephson Current through Semiconductor Nanowire
  with Spin–Orbit Interaction in Magnetic Field}},\ }\href
  {https://doi.org/10.7566/JPSJ.82.054703} {\bibfield  {journal} {\bibinfo
  {journal} {J. Phys. Soc. Jap.}\ }\textbf {\bibinfo {volume} {82}},\ \bibinfo
  {pages} {054703} (\bibinfo {year} {2013})}\BibitemShut {NoStop}%
\bibitem [{\citenamefont {Dolcini}\ \emph {et~al.}(2015)\citenamefont
  {Dolcini}, \citenamefont {Houzet},\ and\ \citenamefont
  {Meyer}}]{Dolcini2015}%
  \BibitemOpen
  \bibfield  {author} {\bibinfo {author} {\bibfnamefont {F.}~\bibnamefont
  {Dolcini}}, \bibinfo {author} {\bibfnamefont {M.}~\bibnamefont {Houzet}},\
  and\ \bibinfo {author} {\bibfnamefont {J.~S.}\ \bibnamefont {Meyer}},\
  }\bibfield  {title} {\bibinfo {title} {{Topological Josephson
  ${\ensuremath{\phi}}_{0}$ junctions}},\ }\href
  {https://doi.org/10.1103/PhysRevB.92.035428} {\bibfield  {journal} {\bibinfo
  {journal} {Phys. Rev. B}\ }\textbf {\bibinfo {volume} {92}},\ \bibinfo
  {pages} {035428} (\bibinfo {year} {2015})}\BibitemShut {NoStop}%
\bibitem [{\citenamefont {Chen}\ \emph {et~al.}(2018)\citenamefont {Chen},
  \citenamefont {He}, \citenamefont {Ali}, \citenamefont {Lee}, \citenamefont
  {Fong},\ and\ \citenamefont {Law}}]{Chen2018}%
  \BibitemOpen
  \bibfield  {author} {\bibinfo {author} {\bibfnamefont {C.-Z.}\ \bibnamefont
  {Chen}}, \bibinfo {author} {\bibfnamefont {J.~J.}\ \bibnamefont {He}},
  \bibinfo {author} {\bibfnamefont {M.~N.}\ \bibnamefont {Ali}}, \bibinfo
  {author} {\bibfnamefont {G.-H.}\ \bibnamefont {Lee}}, \bibinfo {author}
  {\bibfnamefont {K.~C.}\ \bibnamefont {Fong}},\ and\ \bibinfo {author}
  {\bibfnamefont {K.~T.}\ \bibnamefont {Law}},\ }\bibfield  {title} {\bibinfo
  {title} {{Asymmetric Josephson effect in inversion symmetry breaking
  topological materials}},\ }\href {https://doi.org/10.1103/PhysRevB.98.075430}
  {\bibfield  {journal} {\bibinfo  {journal} {Phys. Rev. B}\ }\textbf {\bibinfo
  {volume} {98}},\ \bibinfo {pages} {075430} (\bibinfo {year}
  {2018})}\BibitemShut {NoStop}%
\bibitem [{\citenamefont {Davydova}\ \emph {et~al.}(2022)\citenamefont
  {Davydova}, \citenamefont {Prembabu},\ and\ \citenamefont
  {Fu}}]{Davydova2022}%
  \BibitemOpen
  \bibfield  {author} {\bibinfo {author} {\bibfnamefont {M.}~\bibnamefont
  {Davydova}}, \bibinfo {author} {\bibfnamefont {S.}~\bibnamefont {Prembabu}},\
  and\ \bibinfo {author} {\bibfnamefont {L.}~\bibnamefont {Fu}},\ }\bibfield
  {title} {\bibinfo {title} {{Universal Josephson diode effect}},\ }\href
  {https://doi.org/10.1126/sciadv.abo0309} {\bibfield  {journal} {\bibinfo
  {journal} {Science Adv.}\ }\textbf {\bibinfo {volume} {8}},\ \bibinfo {pages}
  {eabo0309} (\bibinfo {year} {2022})}\BibitemShut {NoStop}%
\bibitem [{\citenamefont {Souto}\ \emph {et~al.}()\citenamefont {Souto},
  \citenamefont {Leijnse},\ and\ \citenamefont {Schrade}}]{Souto2022}%
  \BibitemOpen
  \bibfield  {author} {\bibinfo {author} {\bibfnamefont {R.~S.}\ \bibnamefont
  {Souto}}, \bibinfo {author} {\bibfnamefont {M.}~\bibnamefont {Leijnse}},\
  and\ \bibinfo {author} {\bibfnamefont {C.}~\bibnamefont {Schrade}},\ }\href
  {https://doi.org/10.48550/ARXIV.2205.04469} {\bibinfo {title} {{The Josephson
  diode effect in supercurrent interferometers}}},\ \bibinfo {howpublished}
  {arXiv:2205.04469 (2022)}\BibitemShut {NoStop}%
\bibitem [{\citenamefont {Misaki}\ and\ \citenamefont
  {Nagaosa}(2021)}]{Misaki2021}%
  \BibitemOpen
  \bibfield  {author} {\bibinfo {author} {\bibfnamefont {K.}~\bibnamefont
  {Misaki}}\ and\ \bibinfo {author} {\bibfnamefont {N.}~\bibnamefont
  {Nagaosa}},\ }\bibfield  {title} {\bibinfo {title} {{Theory of the
  nonreciprocal Josephson effect}},\ }\href
  {https://doi.org/10.1103/PhysRevB.103.245302} {\bibfield  {journal} {\bibinfo
   {journal} {Phys. Rev. B}\ }\textbf {\bibinfo {volume} {103}},\ \bibinfo
  {pages} {245302} (\bibinfo {year} {2021})}\BibitemShut {NoStop}%
\bibitem [{\citenamefont {Ruby}\ \emph {et~al.}(2016)\citenamefont {Ruby},
  \citenamefont {Peng}, \citenamefont {von Oppen}, \citenamefont {Heinrich},\
  and\ \citenamefont {Franke}}]{Ruby2016}%
  \BibitemOpen
  \bibfield  {author} {\bibinfo {author} {\bibfnamefont {M.}~\bibnamefont
  {Ruby}}, \bibinfo {author} {\bibfnamefont {Y.}~\bibnamefont {Peng}}, \bibinfo
  {author} {\bibfnamefont {F.}~\bibnamefont {von Oppen}}, \bibinfo {author}
  {\bibfnamefont {B.~W.}\ \bibnamefont {Heinrich}},\ and\ \bibinfo {author}
  {\bibfnamefont {K.~J.}\ \bibnamefont {Franke}},\ }\bibfield  {title}
  {\bibinfo {title} {{Orbital Picture of Yu-Shiba-Rusinov Multiplets}},\ }\href
  {https://doi.org/10.1103/PhysRevLett.117.186801} {\bibfield  {journal}
  {\bibinfo  {journal} {Phys. Rev. Lett.}\ }\textbf {\bibinfo {volume} {117}},\
  \bibinfo {pages} {186801} (\bibinfo {year} {2016})}\BibitemShut {NoStop}%
\bibitem [{\citenamefont {Ruby}\ \emph
  {et~al.}(2015{\natexlab{a}})\citenamefont {Ruby}, \citenamefont {Pientka},
  \citenamefont {Peng}, \citenamefont {von Oppen}, \citenamefont {Heinrich},\
  and\ \citenamefont {Franke}}]{Ruby2015}%
  \BibitemOpen
  \bibfield  {author} {\bibinfo {author} {\bibfnamefont {M.}~\bibnamefont
  {Ruby}}, \bibinfo {author} {\bibfnamefont {F.}~\bibnamefont {Pientka}},
  \bibinfo {author} {\bibfnamefont {Y.}~\bibnamefont {Peng}}, \bibinfo {author}
  {\bibfnamefont {F.}~\bibnamefont {von Oppen}}, \bibinfo {author}
  {\bibfnamefont {B.~W.}\ \bibnamefont {Heinrich}},\ and\ \bibinfo {author}
  {\bibfnamefont {K.~J.}\ \bibnamefont {Franke}},\ }\bibfield  {title}
  {\bibinfo {title} {{Tunneling Processes into Localized Subgap States in
  Superconductors}},\ }\href {https://doi.org/10.1103/PhysRevLett.115.087001}
  {\bibfield  {journal} {\bibinfo  {journal} {Phys. Rev. Lett.}\ }\textbf
  {\bibinfo {volume} {115}},\ \bibinfo {pages} {087001} (\bibinfo {year}
  {2015}{\natexlab{a}})}\BibitemShut {NoStop}%
\bibitem [{\citenamefont {Yazdani}\ \emph {et~al.}(1997)\citenamefont
  {Yazdani}, \citenamefont {Jones}, \citenamefont {Lutz}, \citenamefont
  {Crommie},\ and\ \citenamefont {Eigler}}]{Yazdani1997}%
  \BibitemOpen
  \bibfield  {author} {\bibinfo {author} {\bibfnamefont {A.}~\bibnamefont
  {Yazdani}}, \bibinfo {author} {\bibfnamefont {B.~A.}\ \bibnamefont {Jones}},
  \bibinfo {author} {\bibfnamefont {C.~P.}\ \bibnamefont {Lutz}}, \bibinfo
  {author} {\bibfnamefont {M.~F.}\ \bibnamefont {Crommie}},\ and\ \bibinfo
  {author} {\bibfnamefont {D.~M.}\ \bibnamefont {Eigler}},\ }\bibfield  {title}
  {\bibinfo {title} {{Probing the Local Effects of Magnetic Impurities on
  Superconductivity}},\ }\href {https://doi.org/10.1126/science.275.5307.1767}
  {\bibfield  {journal} {\bibinfo  {journal} {Science}\ }\textbf {\bibinfo
  {volume} {275}},\ \bibinfo {pages} {1767} (\bibinfo {year}
  {1997})}\BibitemShut {NoStop}%
\bibitem [{\citenamefont {Choi}\ \emph {et~al.}(2017)\citenamefont {Choi},
  \citenamefont {Rubio-Verd{\'{u}}}, \citenamefont {{De Bruijckere}},
  \citenamefont {Ugeda}, \citenamefont {Lorente},\ and\ \citenamefont
  {Pascual}}]{Choi2017}%
  \BibitemOpen
  \bibfield  {author} {\bibinfo {author} {\bibfnamefont {D.-J.}\ \bibnamefont
  {Choi}}, \bibinfo {author} {\bibfnamefont {C.}~\bibnamefont
  {Rubio-Verd{\'{u}}}}, \bibinfo {author} {\bibfnamefont {J.}~\bibnamefont {{De
  Bruijckere}}}, \bibinfo {author} {\bibfnamefont {M.~M.}\ \bibnamefont
  {Ugeda}}, \bibinfo {author} {\bibfnamefont {N.}~\bibnamefont {Lorente}},\
  and\ \bibinfo {author} {\bibfnamefont {J.~I.}\ \bibnamefont {Pascual}},\
  }\bibfield  {title} {\bibinfo {title} {{Mapping the orbital structure of
  impurity bound states in a superconductor}},\ }\href
  {https://doi.org/10.1038/ncomms15175} {\bibfield  {journal} {\bibinfo
  {journal} {Nature Commun.}\ }\textbf {\bibinfo {volume} {8}},\ \bibinfo
  {pages} {15175} (\bibinfo {year} {2017})}\BibitemShut {NoStop}%
\bibitem [{\citenamefont {Farinacci}\ \emph {et~al.}(2018)\citenamefont
  {Farinacci}, \citenamefont {Ahmadi}, \citenamefont {Reecht}, \citenamefont
  {Ruby}, \citenamefont {Bogdanoff}, \citenamefont {Peters}, \citenamefont
  {Heinrich}, \citenamefont {von Oppen},\ and\ \citenamefont
  {Franke}}]{Farinacci2018}%
  \BibitemOpen
  \bibfield  {author} {\bibinfo {author} {\bibfnamefont {L.}~\bibnamefont
  {Farinacci}}, \bibinfo {author} {\bibfnamefont {G.}~\bibnamefont {Ahmadi}},
  \bibinfo {author} {\bibfnamefont {G.}~\bibnamefont {Reecht}}, \bibinfo
  {author} {\bibfnamefont {M.}~\bibnamefont {Ruby}}, \bibinfo {author}
  {\bibfnamefont {N.}~\bibnamefont {Bogdanoff}}, \bibinfo {author}
  {\bibfnamefont {O.}~\bibnamefont {Peters}}, \bibinfo {author} {\bibfnamefont
  {B.~W.}\ \bibnamefont {Heinrich}}, \bibinfo {author} {\bibfnamefont
  {F.}~\bibnamefont {von Oppen}},\ and\ \bibinfo {author} {\bibfnamefont
  {K.~J.}\ \bibnamefont {Franke}},\ }\bibfield  {title} {\bibinfo {title}
  {{Tuning the Coupling of an Individual Magnetic Impurity to a Superconductor:
  Quantum Phase Transition and Transport}},\ }\href
  {https://doi.org/10.1103/PhysRevLett.121.196803} {\bibfield  {journal}
  {\bibinfo  {journal} {Phys. Rev. Lett.}\ }\textbf {\bibinfo {volume} {121}},\
  \bibinfo {pages} {196803} (\bibinfo {year} {2018})}\BibitemShut {NoStop}%
\bibitem [{\citenamefont {Ruby}\ \emph
  {et~al.}(2015{\natexlab{b}})\citenamefont {Ruby}, \citenamefont {Heinrich},
  \citenamefont {Pascual},\ and\ \citenamefont {Franke}}]{RubyPb15}%
  \BibitemOpen
  \bibfield  {author} {\bibinfo {author} {\bibfnamefont {M.}~\bibnamefont
  {Ruby}}, \bibinfo {author} {\bibfnamefont {B.~W.}\ \bibnamefont {Heinrich}},
  \bibinfo {author} {\bibfnamefont {J.~I.}\ \bibnamefont {Pascual}},\ and\
  \bibinfo {author} {\bibfnamefont {K.~J.}\ \bibnamefont {Franke}},\ }\bibfield
   {title} {\bibinfo {title} {{Experimental Demonstration of a Two-Band
  Superconducting State for Lead Using Scanning Tunneling Spectroscopy}},\
  }\href {https://doi.org/10.1103/PhysRevLett.114.157001} {\bibfield  {journal}
  {\bibinfo  {journal} {Phys. Rev. Lett.}\ }\textbf {\bibinfo {volume} {114}},\
  \bibinfo {pages} {157001} (\bibinfo {year} {2015}{\natexlab{b}})}\BibitemShut
  {NoStop}%
\end{thebibliography}

\begin{thebibliography}{1}
\expandafter\ifx\csname url\endcsname\relax
  \def\url#1{\texttt{#1}}\fi
\expandafter\ifx\csname urlprefix\endcsname\relax\def\urlprefix{URL }\fi
\providecommand{\bibinfo}[2]{#2}
\providecommand{\eprint}[2][]{\url{#2}}

\bibitem{SAmbegaokar1969}
\bibinfo{author}{Ambegaokar, V.} \& \bibinfo{author}{Halperin, B.~I.}
\newblock \bibinfo{title}{{Voltage Due to Thermal Noise in the dc Josephson
  Effect}}.
\newblock \emph{\bibinfo{journal}{Phys. Rev. Lett.}}
  \textbf{\bibinfo{volume}{22}}, \bibinfo{pages}{1364--1366}
  (\bibinfo{year}{1969}).
\newblock \urlprefix\url{https://link.aps.org/doi/10.1103/PhysRevLett.22.1364}.

\bibitem{SIvanchenko1969}
\bibinfo{author}{{Ivanchenko}, Y.~M.} \& \bibinfo{author}{{Zil'berman}, L.~A.}
\newblock \bibinfo{title}{{The Josephson Effect in Small Tunnel Contacts}}.
\newblock \emph{\bibinfo{journal}{Sov. JETP}} \textbf{\bibinfo{volume}{28}},
  \bibinfo{pages}{1272} (\bibinfo{year}{1969}).

\bibitem{SKautz1990}
\bibinfo{author}{Kautz, R.~L.} \& \bibinfo{author}{Martinis, J.~M.}
\newblock \bibinfo{title}{{Noise-affected I-V curves in small hysteretic
  Josephson junctions}}.
\newblock \emph{\bibinfo{journal}{Phys. Rev. B}} \textbf{\bibinfo{volume}{42}},
  \bibinfo{pages}{9903--9937} (\bibinfo{year}{1990}).
\newblock \urlprefix\url{https://link.aps.org/doi/10.1103/PhysRevB.42.9903}.

\bibitem{SGrabert1999}
\bibinfo{author}{Grabert, H.} \& \bibinfo{author}{Ingold, G.-L.}
\newblock \bibinfo{title}{{Mesoscopic Josephson effect}}.
\newblock \emph{\bibinfo{journal}{Superlattices Microstruct.}}
  \textbf{\bibinfo{volume}{25}}, \bibinfo{pages}{915--923}
  (\bibinfo{year}{1999}).

\end{thebibliography}

\end{document}